\documentclass[a4paper,fleqn]{cas-sc}

\usepackage[numbers,sort&compress]{natbib}
\usepackage{graphicx}
\usepackage{amsmath}
\usepackage{amssymb}
\usepackage{subcaption}
\usepackage{lineno}

\newcommand{\kmwe}{\mathrm{km.w.e.}}
\newcommand{\pe}{\mathrm{PE}}

\ExplSyntaxOn
\cs_gset_eq:NN \vbox_unpack_clear:N \vbox_unpack_drop:N
\ExplSyntaxOff

\begin{document}
\let\WriteBookmarks\relax

\shorttitle{Candidate Earth-skimming neutrinos at the HAWC observatory} \shortauthors{H. Le\'on~Vargas and A. Sandoval}

\title[mode = title]{Anomalous horizontal track-like events at the HAWC observatory: candidate neutrino-induced charged leptons from the Pico de Orizaba volcano}

\author[1]{Hermes Le\'on~Vargas}[orcid=0000-0001-5516-4975] \cormark[1] \ead{hleonvar@fisica.unam.mx}

\author[1]{Andr\'es Sandoval}

\affiliation[1]{organization={Instituto de F\'isica, Universidad Nacional Aut\'onoma de M\'exico}, addressline={Circuito de la Investigaci\'on Cient\'ifica s/n, Ciudad Universitaria}, postcode={04510}, city={Ciudad de M\'exico}, country={M\'exico}}

\cortext[1]{Corresponding author.}

\begin{abstract}
Two track-like events with exceptionally large charge deposits, pointing back to the region of maximum overburden ($>18~\kmwe$) of the Pico de Orizaba volcano, were reported by the HAWC Collaboration in a search for Earth-skimming neutrinos. Using exclusively published information, including the published Monte Carlo charge distributions and the scattered-muon background model, we show that these events are kinematically inconsistent with all identified backgrounds. A conservative energy-conservation bound places the in-tank energy deposit of the candidates above $\approx$190--210~GeV, with a worst-case minimum of 122--136~GeV, in either case exceeding the 100~GeV maximum of the scattered-muon background model. Calibration systematics derived from the published simulations act unidirectionally, pushing the implied deposits into the multi-TeV regime. Direct atmospheric muons require $E_\mu \gtrsim 0.67$~PeV to penetrate the overburden and contribute $4.5\times10^{-5}$ expected events ($\lesssim 10^{-3}$ under the largest possible prompt-charm flux); collinear muon bundles are excluded by five independent arguments. Conversely, the neutrino hypothesis predicts identifiable events precisely in the observed charge window ($10^{3}$--$10^{4}$ photoelectrons) and in the highest-overburden analysis region, both observed. The two track-like events imply a rate that exceeds the conventional atmospheric neutrino flux expectation at the 2.1--2.5$\sigma$ level. The charges of the observed candidates require an emerging lepton above about 20~TeV; the published calibration does not bound the energy from above. These would constitute the first neutrino candidates detected in Mexico and the first obtained with the Earth-skimming volcano technique.
\end{abstract}


\maketitle

\section{Introduction} \label{sec:intro}

Anomalous near-horizontal events have a long history in underground and surface particle physics. The Kolar Gold Fields (KGF) experiments reported anomalous multi-track events with large opening angles and vertices in the air or in thin detector material. These events could not be explained by the known atmospheric or neutrino-induced backgrounds \cite{KGF1971,KGF1975}, a puzzle that remains open five decades later. The same program had earlier yielded, together with the Case--Witwatersrand--Irvine experiment, the first detections of atmospheric neutrinos, obtained precisely as near-horizontal muons deep underground \cite{Achar1965,Reines1965}. Other landmark underground neutrino programs in this regime include the Fr\'ejus, MACRO, and Soudan~2 experiments \cite{Frejus1996,MACRO2003,Soudan2_1999}. The LVD experiment at Gran Sasso measured the neutrino-induced horizontal muon intensity over five decades of intensity \cite{LVD1995}, providing the quantitative anchor against which modern surface-detector searches such as HAWC are compared. More recently, the IceCube Collaboration reported a statistically significant excess of data over simulation for reconstructed tracks in the near-horizontal band ($90^\circ$--$97^\circ$ zenith angle) of its 40-string configuration \cite{IC40}, similar to an excess previously observed with the AMANDA-II detector \cite{AMANDA2010}. Dedicated checks showed that the excess decreases neither with depth nor with tightened quality cuts (a behavior consistent with muons from atmospheric neutrino interactions), although residual misreconstructed atmospheric muons could not be excluded, due in part to uncertainties in the modeling of photon propagation through the layered ice and in the cosmic-ray shower simulation. Since its origin could not be verified, the near-horizontal band was removed from the analysis \cite{IC40}. These cases illustrate a structural limitation of in-ice and underground neutrino telescopes: the region around the horizon, where the Earth becomes transparent to the highest-energy neutrinos, is also the most contaminated by misreconstructed muon backgrounds, and is therefore routinely discarded. That the most energetic neutrino recorded to date, KM3-230213A, arrived from less than a degree above the horizon \cite{KM3NeT2025} sharpens the point: the band that these detectors find hardest to control is also the one in which the highest-energy events appear, and in which consistency between experiments has proved hardest to establish (Section~\ref{sec:discussion}). The 2022 Snowmass Neutrino Frontier reports highlight this challenge, noting that next-generation detectors must overcome these backgrounds to resolve neutrinos at and above PeV energies \cite{SnowmassNF10,SnowmassHE}. The Earth-skimming method, with surface detectors, inverts this situation: the rock of a mountain suppresses the atmospheric backgrounds precisely in the horizontal direction, converting the problematic band into a shielded observation window. Recent proposals such as TAMBO, a deep-valley neutrino observatory in the Peruvian Andes, aim to exploit this geometry to search for tau neutrinos at supra-PeV energies, with projected sensitivity to the diffuse flux between $\sim$3~PeV and 1~EeV exceeding that of all present-day observatories \cite{TAMBO2026}.

The Earth-skimming technique \cite{Fargion1999,Feng2002} exploits a large mass of rock (a mountain or a chord through the Earth) as the target for charged-current neutrino--nucleon interactions, detecting the emerging charged lepton. This method has been employed by experiments such as Pierre Auger, MAGIC, and Ashra to set upper limits on the ultrahigh-energy neutrino flux \cite{Auger2009,MAGIC2018,Ashra2013}. Its implementation with the High Altitude Water Cherenkov (HAWC) observatory, using the Pico de Orizaba volcano as a neutrino target and as a background shield, was proposed in Ref.~\cite{LeonVargas2017}. This proposal leverages HAWC's modular design, which allows the 22\,000~m$^2$ array to operate as a high-precision horizontal particle tracker. It predicted that neutrino-induced leptons or their collimated decay products would deposit thousands to tens of thousands of photoelectrons (PE) along track-like topologies, one to two orders of magnitude above the typical atmospheric-muon signals of $\approx 30$~PE \cite{Albert2022,Smith2015}. At PeV energies, tau leptons emerging from the volcano produce highly collimated air showers (opening angles $<0.2^\circ$) that appear as clean tracks across HAWC's large Water Cherenkov Detectors (WCDs) \cite{LeonVargas2017}. A dedicated tracking analysis of six months of HAWC data subsequently identified two track-like events with average charge deposits per detector of 1561.7 and 1744.8~PE, pointing back to the base of the volcano where the overburden exceeds $18~\kmwe$ \cite{ICRC2019}. Finally, a refereed background study \cite{Albert2022} established that the population of reconstructed horizontal tracks is quantitatively explained by low-energy atmospheric muons ($2 \le E_\mu \le 100$~GeV, most probable energy $\approx 4$~GeV) scattered into the horizontal acceptance. This scattering background, whose deflection angle scales inversely with muon momentum, remains the primary challenge for surface arrays.

The chronology matters for the logical structure of the present work: the charge regime of the signal was predicted (2017) before the two events were observed (2019), and the background model published afterwards (2022) demonstrates that this regime is unreachable by the background. In this paper we close the inference chain. Using exclusively published information, we show that: (i) both candidates are kinematically excluded as scattered muons by energy conservation; (ii) direct muons and collinear muon bundles are negligible or excluded; (iii) the neutrino hypothesis is not merely allowed but predictive, placing identifiable events exactly in the observed charge window and in the observed cells, above the minimum lepton energy that the candidate charges require. We discuss the mild tension in the absolute rate and its natural interpretation.

All the numerical inputs used in this analysis are published values, compiled in Table~\ref{tab:inputs}; no unpublished HAWC result is employed. The event displays of Fig.~\ref{fig:displays} have been redrawn from the corresponding published figures of Ref.~\cite{ICRC2019}.

\section{Published inputs} \label{sec:inputs}

The search of Ref.~\cite{ICRC2019} divides the solid angle subtended by the volcano into twelve rectangular cells, $6^\circ$ wide in azimuth and $4^\circ$ high in elevation, labelled A to L, and each candidate is identified throughout this work by the cell that contains its reconstructed direction. The cells tile the silhouette of the volcano in a lower row centered at $2^\circ$ elevation and an upper row centered at $6^\circ$. The analysis region used below comprises the three azimuth bins closest to the summit, centered at $309^\circ$, $315^\circ$ and $321^\circ$, taken in both rows: cells C, D and E in the lower row and I, J and K in the upper one. It is therefore fixed by the topography alone, together with the requirement that the two rows be paired at common azimuth. Fig.~\ref{fig:cells} shows the partition superimposed on the silhouette of the volcano as seen from HAWC.

Table~\ref{tab:inputs} lists every experimental input used in this work, each with its published source. 

\begin{table}[pos=htbp]
\centering \caption{Published inputs used in this analysis. $\theta_{\rm Rec}$ denotes the reconstructed elevation angle above the horizon. Charges are quoted as $\langle q \rangle$, a mean charge registered in a WCD, with the subscript identifying the average that is taken: $\langle q \rangle_{\rm pixel}$ is the average over the pixels (WCDs) traversed by a single track. The topology cuts HA and MHA are defined in \citep{ICRC2019} and are equivalent to those used in \citep{Albert2022} to select isolated tracks within the air shower triggered sample. The acceptance corresponds to the one used in \citep{Albert2022}, i.e. restricted to the most efficient cells presented in Figure \ref{fig:cells}.} \label{tab:inputs} \small
\begin{tabular}{l l l}
\hline
Quantity & Value & Source \\
\hline
Vertical-muon mean charge & $\langle q \rangle \approx 30~\pe$ & \cite{Albert2022,Smith2015} \\
Candidate 1 (Run 7136, cell C) & $\langle q \rangle_{\rm pixel} = 1561.7~\pe$ & \cite{ICRC2019} \\
Candidate 2 (Run 7659, cell E) & $\langle q \rangle_{\rm pixel} = 1744.8~\pe$ & \cite{ICRC2019} \\
Minimum track length & $\rm TL \ge 4$ pixels (WCDs) & \cite{ICRC2019,Albert2022} \\
MC mean charges, 10~GeV--100~TeV & $\langle q \rangle_{\rm MC} = 26.5$--$385.6~\pe$ (six cells) & \cite{ICRC2019} \\
Background population mean charge & $\langle q \rangle \approx 65$--$84~\pe$ (six cells) & \cite{ICRC2019} \\
Scattered-muon model range & $2 \le E_\mu \le 100$~GeV & \cite{Albert2022} \\
Most probable scattered energy & $E_\mu \approx 4$~GeV & \cite{Albert2022} \\
Overburden of cells C, E (lower row) & $>18~\kmwe$ & \cite{ICRC2019} \\
Elevation resolution ($\theta_{\rm Rec}<2^\circ$, high-efficiency azimuth bins) & $\sigma_\theta = 0.7^\circ$ & \cite{Albert2022} \\
Topology cuts (HA, MHA) & 99.93\% rejection, $<1\%$ FP & \cite{Albert2022} \\
Acceptance (full region, low $E$) & $A\Omega = 5.1\times10^{-2}~{\rm m^2 sr}$ & \cite{Albert2022} \\
Horizontal atmospheric $\nu_\mu + \bar\nu_\mu$ flux &
  \begin{tabular}[t]{@{}l@{}}
  $\Phi_\nu \approx 10^{-10}\,(E/{\rm TeV})^{-3.7}$ \\
  ${\rm GeV^{-1}cm^{-2}s^{-1}sr^{-1}}$
  \end{tabular} & \cite{Volkova1980,GaisserHonda2002} \\
Astrophysical $\nu$ flux (per flavor) &
  \begin{tabular}[t]{@{}l@{}}
  $\Phi_\nu = 1.4\times10^{-18}\,(E_\nu/100~{\rm TeV})^{-2.5}$ \\
  ${\rm GeV^{-1}cm^{-2}s^{-1}sr^{-1}}$
  \end{tabular} & \cite{IceCubeTG2022,IceCubeST2024} \\
Live time & $T \approx 181$~days & \cite{ICRC2019} \\
WCD geometry & $\varnothing\,7.3$~m $\times$ 4.5~m depth & \cite{HAWCDetector2023} \\
Volcano rock density (standard rock) & $\rho = 2.65~{\rm g/cm^3}$ & \cite{LeonVargas2017,Lohmann1985,Groom2001} \\
Electronics dynamic range & up to thousands of PE & \cite{Albert2022} \\
\hline
\end{tabular}
\end{table}

One additional property of the candidates, extractable from the published event displays, is worth recording: both tracks exhibit sustained large charge deposits along their full length, rather than a single bright PMT.

\begin{figure}[pos=htbp]
\centering \includegraphics[width=0.95\linewidth]{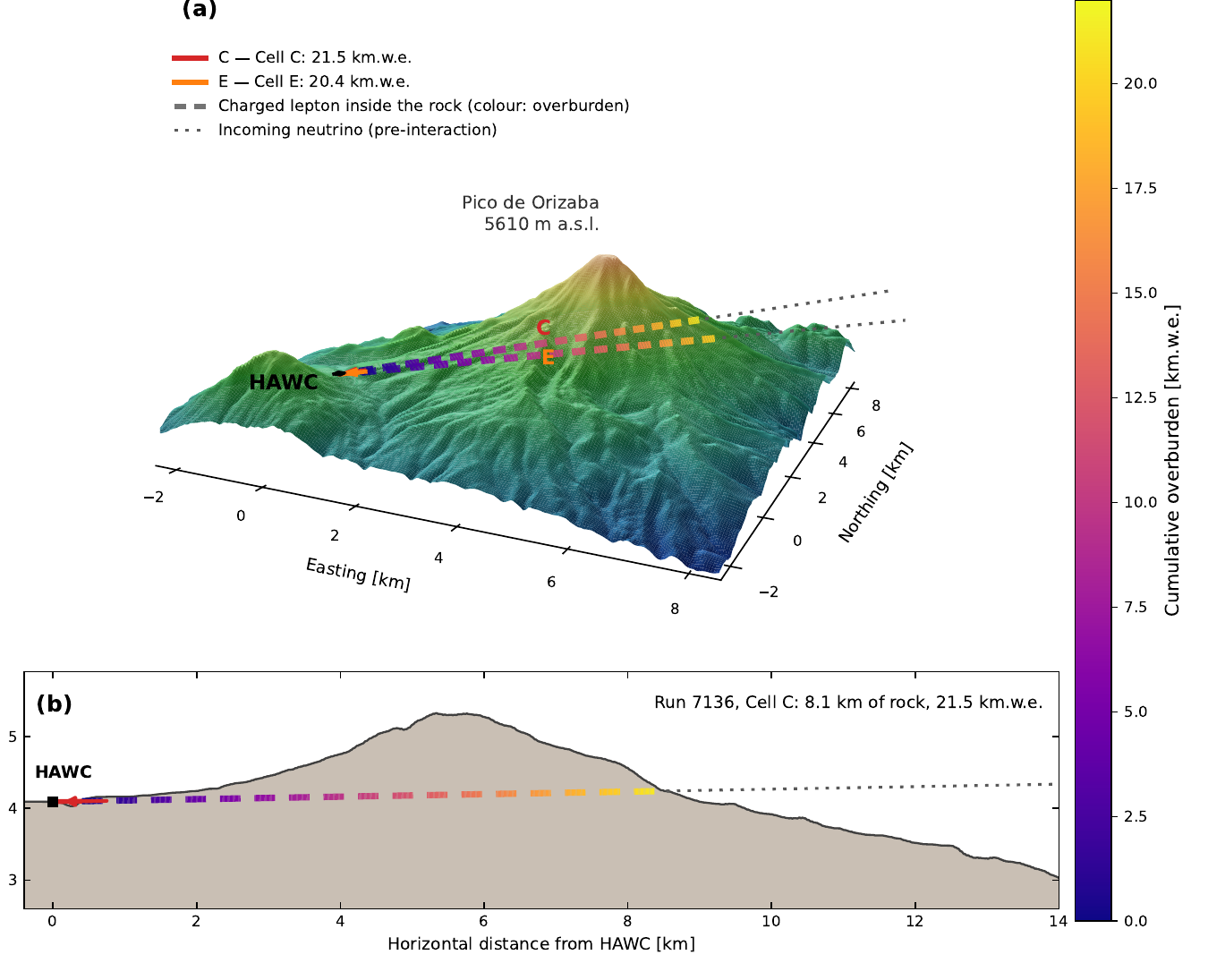} \caption{Topography of the Pico de Orizaba volcano from the INEGI digital elevation model. (a) The two rays are traced back from the HAWC position along the candidate track directions measured in Section~\ref{sec:inputs}, at $1^\circ$ elevation. The dotted gray segment is the incoming neutrino before its interaction, the dashed color-coded segment the charged lepton inside the rock, and the solid colored segment the lepton in air; the color scale gives the cumulative overburden in $\kmwe$, which reaches 21.5 and 20.4~$\kmwe$ for cells C and E respectively. These are single-direction values and therefore exceed the cell-averaged figures quoted in the text, which average over the full solid angle of each cell. (b) Vertical section along the azimuth of the candidate of cell C,  in which the trajectory is seen to cross 8.1~km of rock.} \label{fig:hero}
\end{figure}

\begin{figure}[pos=htbp]
\centering \includegraphics[width=0.95\linewidth]{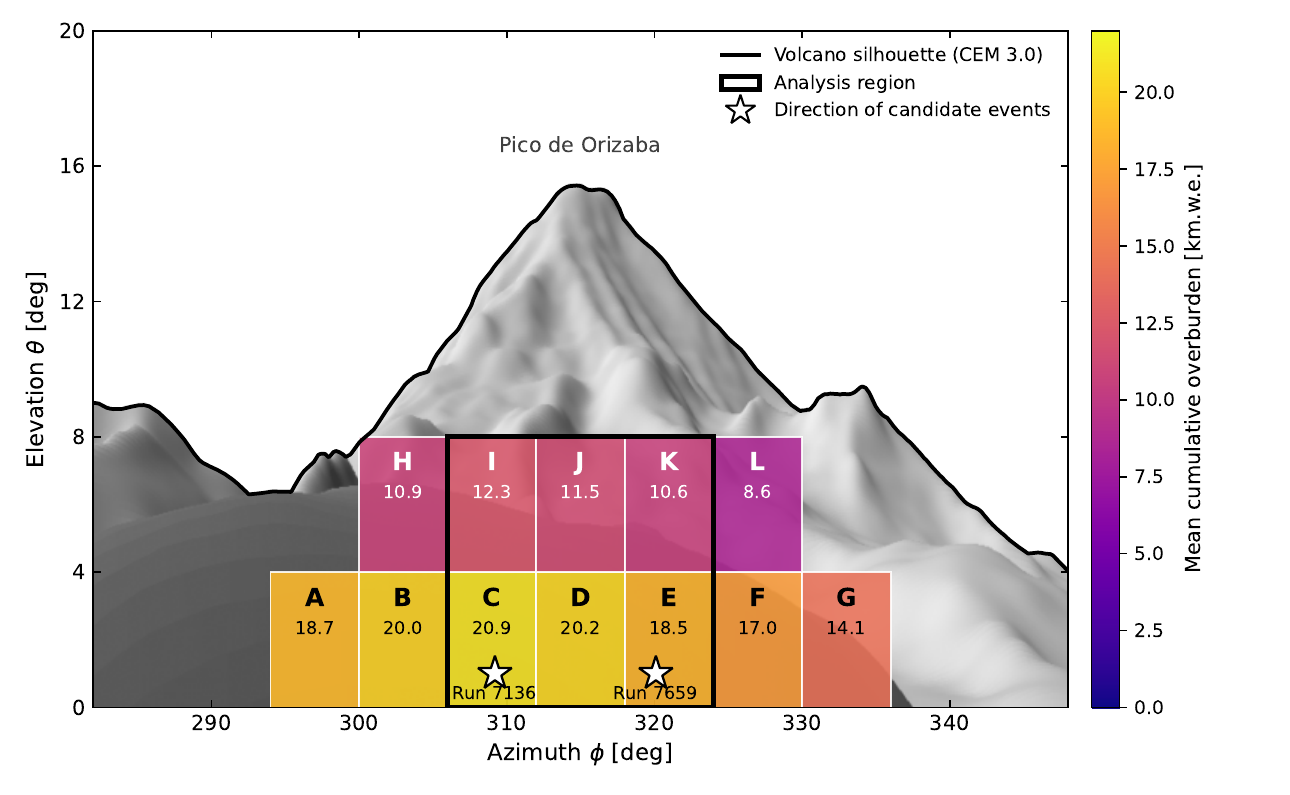} \caption{Analysis cells of Ref.~\cite{ICRC2019} on the silhouette of the Pico de Orizaba as seen from the center of the HAWC array. The relief is obtained by ray tracing the INEGI digital elevation model (CEM~3.0) \cite{INEGI} from the HAWC position, and the azimuth follows the convention of Ref.~\cite{Albert2022}, measured from East and clockwise. Each cell is shaded by the cumulative overburden averaged over its solid angle, from the same ray tracing; the value in $\kmwe$ is printed inside each cell. The heavy outline marks the six-cell analysis region and the stars mark the reconstructed directions of the two observed candidates. The lower row, which contains both candidates, is shielded by roughly twice the mass of the upper row, and part of cell H looks past the edge of the volcano. The smooth dark mass in the left foreground is a lava structure a few hundred metres from the array.} \label{fig:cells}
\end{figure}

\begin{figure}[pos=htbp]
\centering \includegraphics[width=\textwidth]{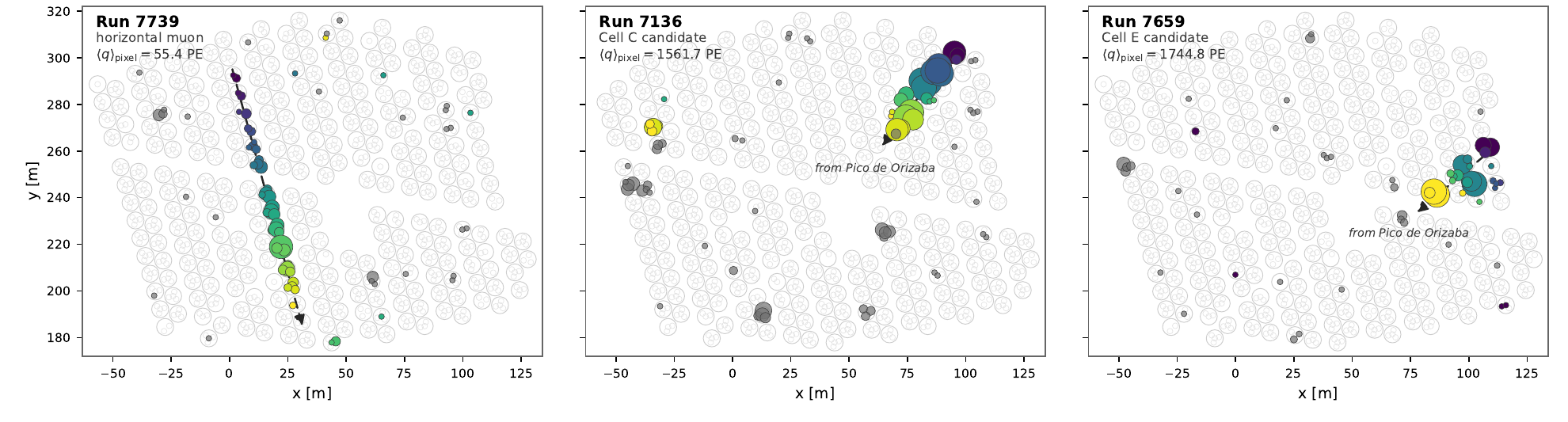} \caption{Event displays published in Ref.~\cite{ICRC2019}, redrawn from the original figures. Left: the horizontal muon of Run 7739, which serves as a reference from the same instrument. Center and right: the two candidates. The three panels share axes, aspect ratio and marker scale, so that marker sizes are directly comparable. Marker size gives the deposited charge, on the scale of the original displays, whose minimum is the PMT radius and whose maximum is $1.5$ times the tank radius. Marker color encodes the relative arrival time of each PMT hit; the color range is set independently in each panel to the interval spanned by that panel's track, no absolute time scale is quoted, and hits arriving outside that interval are shown in gray. The dashed arrows indicate the approximate direction of the track defined by the highest-charge hits, and are a visual guide only. The isolated low-charge hits correspond to the vertical atmospheric muon noise contained in the same trigger window.} \label{fig:displays}
\end{figure}

We independently validated the published overburden values, referred to throughout this work as the cumulative overburden and equivalent to the line-of-sight mass of Ref.~\cite{Albert2022}, using the public INEGI digital elevation model (CEM~3.0, 15~m resolution) \cite{INEGI}, by ray tracing from the HAWC position on a $0.25^\circ$ grid and averaging over the solid angle of each cell rather than along a single representative direction. The cell-averaged overburdens obtained are 21.0 and 18.6~$\kmwe$ for cells C and E, with a maximum rock path of 8.4~km. Both exceed the $>18~\kmwe$ quoted in Ref.~\cite{ICRC2019}, and the value for cell C agrees with the $20.97~\kmwe$ that Ref.~\cite{Albert2022} reports for the sub-region of its search covering that azimuth (Fig.~\ref{fig:hero}). Both cells are shielded uniformly: no direction within cell C falls below 18.6~$\kmwe$, and only 4~per~cent of cell E falls below 15~$\kmwe$.\footnote{Cell averages are stable, but the extreme values within a cell depend on how its solid angle is sampled: changing the angular grid, the radial step or the interpolation of the elevation model displaces the minimum of a given cell by up to $\approx 1~\kmwe$. The minima quoted here and in Section~\ref{sec:charge} carry that tolerance, which is well below the contrast on which the argument rests.}

The overburden of $>18~\kmwe$ applies to the lower row of cells of Ref.~\cite{ICRC2019}, shown in Fig.~\ref{fig:cells}, at $0^\circ$--$4^\circ$ elevation, which is the row containing both candidates, and it should not be read as a property of the search region as a whole: the upper row of the same region, at $4^\circ$--$8^\circ$, is shielded by roughly half as much, 10.5 to 12.1~$\kmwe$. Every statement about $>18~\kmwe$ in what follows refers to the lower row, and in practice to the two cells that contain the candidates.

Three checks establish that this geometry and our reading of the published event displays reproduce quantities that we did not use as input. First, the azimuth convention of Ref.~\cite{Albert2022}, measured from East and clockwise, places the modeled summit at $\phi = 315^\circ$, while the elevation profile computed from the digital elevation model peaks at $\phi = 314.5^\circ$. Second, fitting a straight line to the centers of the tanks carrying signal in the published display of Run 7739 returns $\phi = 254.5^\circ$, against the $254.6 \pm 2.8^\circ$ of Table~1 of Ref.~\cite{ICRC2019}. Third, the same procedure applied to the two candidates returns $\phi = 309.2^\circ$ and $320.1^\circ$, against the centers of cells C and E at $309^\circ$ and $321^\circ$; those cell assignments are stated in Ref.~\cite{ICRC2019} and enter nowhere in the fit.

Two entries of Table~\ref{tab:inputs} deserve explicit justification. The vertical-muon anchor $\langle q \rangle \approx 30~\pe$ is the standard single-muon response of a HAWC WCD. It is quoted as such by the background study \cite{Albert2022}, which attributes it to Ref.~\cite{Smith2015}, where charges above $30~\pe$ far from the shower core serve as the muon-identification criterion, and it is the isolated-particle reference used by both that study and the neutrino-detection proposal \cite{LeonVargas2017}. The per-pixel deposits of the candidates exceed it by factors of 52 and 58. The rock density $\rho = 2.65~{\rm g/cm^3}$ is the standard-rock convention of underground particle physics, tracing back to Ref.~\cite{Barrett1952}, extended by Menon and Ramana Murthy \cite{MenonMurthy1967}, and adopted by the muon energy-loss tabulations \cite{Lohmann1985,Groom2001}; its use enables direct comparison with underground observatories that follow the same convention. Adopting instead the mean density of andesitic rock, $2.6~{\rm g/cm^3}$ \cite{Albert2022}, would reduce the overburdens by only $\approx 2\%$, without consequence for any bound derived in this work.

\section{Energy calibration and its systematics} \label{sec:calibration}

\subsection{Minimum-ionizing anchor} \label{sec:mip}

The published mean signal of a vertical atmospheric muon crossing a HAWC WCD is $\langle q \rangle \approx 30~\pe$ \cite{Albert2022,Smith2015}. We define the vertical calibration constant $\kappa_v$ as the mean measured charge per unit of energy deposited in the water volume, in units of PE/GeV, for a particle crossing the WCD along its vertical axis. A minimum-ionizing muon deposits $dE/dx \approx 2.0$~MeV/cm in water, i.e.\ $E_{\rm dep} \approx 0.90$~GeV over the 4.5~m depth, yielding\footnote{The official HAWC calibration for electromagnetic particles is $\approx 40~\pe/{\rm GeV}$ \cite{HAWCDetector2023}, a Monte Carlo value for Cherenkov-dense electromagnetic deposits distributed through the volume. Adopting it in place of $\kappa_v$ would move the minima of Eq.~(\ref{eq:bound}) to $\approx$156 and 174~GeV, still above the 100~GeV maximum of the background model.}
\begin{equation}
\kappa_v \;=\; \frac{30~\pe}{0.90~{\rm GeV}} \;\approx\; 33.3~\pe/{\rm GeV}.
\label{eq:kappav}
\end{equation}
For the horizontal geometry we define analogously the horizontal calibration constant $\kappa_h$, the mean charge per unit of deposited energy for trajectories crossing the WCD along a horizontal chord, the geometry of the track-like events considered in this work. The mean chord of a cylinder of diameter $D=7.3$~m is $\langle L \rangle = (\pi/4)D \approx 5.73$~m ($E_{\rm dep} \approx 1.15$~GeV by ionization). The published MC mean charges for 10~GeV muons (essentially MIPs) in cells C and E, 27.4 and 32.7~$\pe$, imply $\kappa_h \approx 24$--$28.5~\pe/{\rm GeV}$. The fact that $\kappa_h < \kappa_v$ is expected from the less favorable light collection of horizontal trajectories with upward-facing PMTs. Throughout this work we use $\kappa = \kappa_v = 33.3~\pe/{\rm GeV}$ for energy bounds, because it is the largest calibration value and therefore yields the most conservative (lowest) energy estimates.

\subsection{Effective conversion factor versus signal brightness} \label{sec:keff}

The published MC mean charges for mono-energetic muons at 10~GeV, 100~GeV, 1~TeV, 5~TeV and 100~TeV \cite{ICRC2019} allow a direct determination of the effective conversion factor in each brightness regime, since they were produced with the full detector simulation (light collection, PMT response, ToT digitization, and selection cuts included). We define the effective conversion factor $\kappa_{\rm eff}(E)$ as the incremental charge registered per unit of additional energy deposited, relative to the 10~GeV (MIP) baseline, that is, the differential response of the detector in the brightness regime sampled by muons of energy $E$:
\begin{equation}
\kappa_{\rm eff}(E) \;=\;
   \frac{\Delta \langle q \rangle_{\rm MC}(E)}{\Delta E_{\rm dep}(E)},
\qquad
\begin{aligned}
\Delta \langle q \rangle_{\rm MC}(E) &\;\equiv\; \langle q \rangle_{\rm MC}(E)
   - \langle q \rangle_{\rm MC}(10~{\rm GeV}), \\
\Delta E_{\rm dep}(E) &\;\simeq\; b_w(E)\, \langle L \rangle\, E
   + \delta_{\rm ion},
\end{aligned}
\label{eq:keff}
\end{equation}
where $\langle q \rangle_{\rm MC}(E)$ is the published mean charge per pixel registered for simulated mono-energetic muons of energy $E$, so that $\Delta \langle q \rangle_{\rm MC}(E)$ is the charge registered in excess of the 10~GeV baseline and $\Delta E_{\rm dep}(E)$ the energy deposited in excess of it; $b_w(E)$ is the radiative energy-loss parameter in water \cite{Groom2001} and $\delta_{\rm ion} \approx 0.1$~GeV. The total mean energy loss of muons is conventionally written as $\langle -dE/dx \rangle = a(E) + b(E)\,E$, where $a(E)$ is the electronic (ionization) loss and $b(E)$ scales the radiative contribution \cite{PDG2020,Groom2001}; $b_w(E)$ combines bremsstrahlung, direct $e^+e^-$ pair production, and photonuclear interactions. Radiative processes overtake electronic losses above the muon critical energy, $\approx 1.03$~TeV in liquid water \cite{PDG2020,Groom2001}, so the highest published MC points (and the multi-TeV regime inferred for the candidates in Section~\ref{sec:exclusion}) lie well inside the radiative, cascade-dominated regime. The term $\delta_{\rm ion}$ captures the residual relativistic rise of the ionization loss between 10 and 100~GeV, whose logarithmic growth is moderated at high energy by the density-effect correction $\delta(\beta\gamma)$ (the polarization of the medium) toward the Fermi plateau \cite{PDG2020,Sternheimer1984}. Table~\ref{tab:keff} shows the result for cells C and E; the remaining four published cells reproduce the same pattern within $\pm 20\%$.

\begin{table}[pos=htbp]
\centering \caption{Effective conversion factor derived from the published MC mean charges (cells C and E of Ref.~\cite{ICRC2019}). $\Delta \langle q \rangle_{\rm MC}$ and $\Delta E_{\rm dep}$ are the charge registered and the energy deposited in excess of the 10~GeV baseline, both defined in Eq.~(\ref{eq:keff}).} \label{tab:keff}
\begin{tabular}{l c c c c c}
\hline
Regime & $\Delta E_{\rm dep}$ [GeV] & $\Delta \langle q \rangle_{\rm MC}$ (C) [$\pe$] & $\Delta \langle q \rangle_{\rm MC}$ (E)
[$\pe$] & $\kappa_{\rm eff}$ (C) & $\kappa_{\rm eff}$ (E) \\
\hline
100 GeV & $\sim$0.23 & 8.5 & 11.8 & $\sim$37 & $\sim$51 \\
1 TeV   & $\sim$1.8  & 35.9 & 42.1 & $\sim$20 & $\sim$23 \\
5 TeV   & $\sim$9.4  & 81.7 & 95.1 & $\sim$8.7 & $\sim$10.2 \\
100 TeV & $\sim$200  & 298.9 & 335.9 & $\sim$1.5 & $\sim$1.7 \\
\hline
\end{tabular}
\end{table}

The central result is that $\kappa_{\rm eff}$ decreases monotonically with signal brightness, by more than an order of magnitude across the published range. The physical origins (sub-linearity of the ToT charge estimate at large signals and photon absorption of light produced far from the PMTs) do not need to be disentangled here: their combined magnitude is quantified by the public simulations themselves.

\subsection{Direction of the bias} \label{sec:bias}

\begin{figure}[pos=htbp]
\centering

\includegraphics[width=0.95\linewidth]{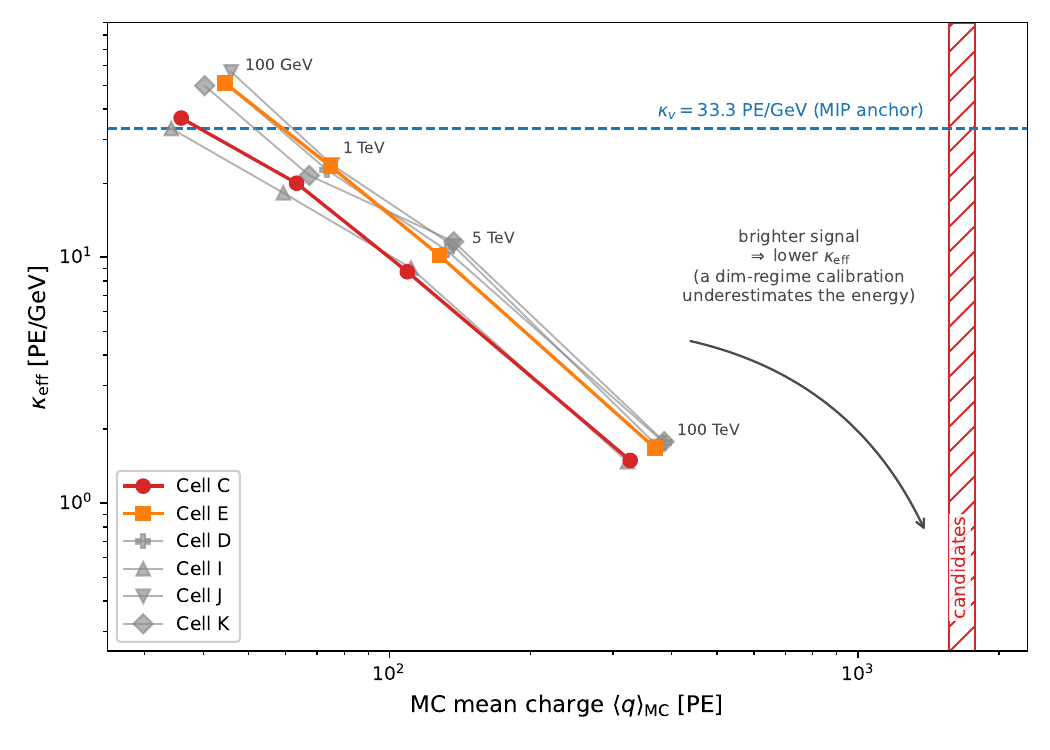}

\caption{Effective conversion factor $\kappa_{\rm eff}$ as a function of the MC mean charge, for the six analysis cells published in Ref.~\cite{ICRC2019}. The monotonic decrease establishes the unidirectional character of the calibration bias: energies estimated with a dim-regime calibration are underestimates.} \label{fig:keff}
\end{figure}

The decisive consequence of Table~\ref{tab:keff}, shown graphically in Fig.~\ref{fig:keff}, is that the calibration bias is unidirectional: converting the charge of a bright deposit with a calibration taken from a dimmer regime can only underestimate the energy. Any residual imperfection in the modeling of PMT saturation operates in the same conservative direction, compressing the measured charge relative to the true light yield. This property is used in Section~\ref{sec:exclusion} to construct a hierarchy of energy bounds.

\section{Kinematic exclusion of the scattered-muon background} \label{sec:exclusion}

\subsection{Charge-space comparison (calibration-free)} \label{sec:chargespace}

The comparison between candidates and background can be performed entirely in charge space, without ever trying to convert charge to energy. The candidates, with 1561.7 and 1744.8~$\pe$ per pixel, exceed by a factor 4.1--5.4 the mean MC charge of 100~TeV muons (322.0--385.6~$\pe$ across the six analysis cells) \cite{ICRC2019}, and by a factor $\approx 19$--27 the mean charge of the background population of those cells ($\approx 65$--$84~\pe$, the means of the published data distributions of Figs.~6--8 of Ref.~\cite{ICRC2019}), whose energies are bounded by $E_\mu \le 100$~GeV in the background model of Ref.~\cite{Albert2022}. The published MC mean charge at 100~GeV in cells C and E is only 35.9--44.5~$\pe$. In other words, the candidates are $\sim$40 times brighter than a muon at the very energy limit of the background model, and $\sim$5 times brighter than an average muon one thousand times more energetic than that limit. Calibration systematics play no role in this comparison.

\subsection{Energy-conservation bound} \label{sec:energybound}

With the minimum published track length ${\rm TL} \ge 4$ pixels, the total measured charge satisfies $Q_{\rm tot} \ge 4\,\langle q \rangle_{\rm pixel}$, i.e.\ 6247~$\pe$ (Run 7136) and 6979~$\pe$ (Run 7659). The energy deposited inside the tanks alone then obeys
\begin{equation}
E_{\rm dep} \;\ge\; \frac{Q_{\rm tot}}{\kappa}
\;\gtrsim\; 190~{\rm GeV} \;\;({\rm Run~7136}),
\qquad 210~{\rm GeV} \;\;({\rm Run~7659}),
\label{eq:bound}
\end{equation}
using the conservative MIP calibration of Eq.~(\ref{eq:kappav}). This bound is robust in four independent, mutually reinforcing directions: (i) TL may exceed 4 (as is the case for both of the candidates); (ii) saturation can only underestimate $Q_{\rm tot}$; (iii) $\kappa=33.3~\pe/{\rm GeV}$ overestimates the light per GeV in horizontal geometry; (iv) the energy lost in the gaps between tanks and the cascade light escaping the sensitive volume are not counted.

The comparison with the background model is then direct, with $E^{\rm scatt}_{\rm max}$ the upper energy limit of the scattered-muon population in the model of Ref.~\cite{Albert2022}:
\begin{equation}
E_{\rm dep} \gtrsim 190~{\rm GeV} \;>\; E^{\rm scatt}_{\rm max} = 100~{\rm GeV}.
\label{eq:exclusion}
\end{equation}
A scattered muon cannot produce either candidate by energy conservation, with a violation factor of 1.9--2.1 even for the most energetic muon in the model. This is not a statistical tail argument: the background model assigns these charges zero probability, and its 100~GeV limit is physical, since the probability of scattering into the horizontal acceptance is inversely proportional to the muon momentum \cite{Albert2022}, so the population capable of reaching these directions vanishes precisely where the energy would become sufficient.

The absolute worst-case bound is obtained by forcing the highest $\kappa_{\rm eff}$ of Table~\ref{tab:keff} ($\approx 51~\pe/{\rm GeV}$, the 100~GeV regime of cell E), physically inapplicable to deposits 40 times brighter but useful as a hard minimum: $E_{\rm dep} \ge 122$ and 136~GeV. Even in this indefensibly pessimistic scenario both candidates remain above the background maximum. Conversely, the calibration consistent with the bright regime actually sampled by the candidates ($\kappa_{\rm eff} \approx 1.5$--$10~\pe/{\rm GeV}$) implies larger deposits. Its weakest member, the 5~TeV-regime value $\kappa_{\rm eff} \approx 10~\pe/{\rm GeV}$, already gives $E_{\rm dep} \gtrsim 610$~GeV. The candidates are, however, $\approx 4.7$ times brighter than the 100~TeV MC point of their own cell, so the applicable entry is the faintest one published, $\kappa_{\rm eff} \approx 1.5$--$1.7~\pe/{\rm GeV}$, which yields $E_{\rm dep} \approx 4.1$~TeV for both candidates after subtracting the four-pixel minimum-ionizing baseline; and since $\kappa_{\rm eff}$ decreases monotonically with brightness (Fig.~\ref{fig:keff}), that value is itself a lower estimate. The full hierarchy is therefore: unassailable minimum $\approx$122~GeV; conservative reference $E_{\rm dep} \approx$ 190~GeV (Eq.~(\ref{eq:bound})); $E_{\rm dep} \gtrsim$ 610~GeV with the 5~TeV-regime calibration; and $\approx$ 4.1~TeV with the faintest calibration of Table~\ref{tab:keff}, all deposited inside the tanks.

Sustaining a deposit of $\sim$47--52~GeV per tank along the track, the conservative reference of Eq.~(\ref{eq:bound}) divided among the four pixels, already requires radiative losses to dominate. In liquid water, electronic and radiative energy losses are equal at the muon critical energy, $E_{\mu c} \approx 1.03$~TeV \cite{Groom2001,PDG2020}; above this threshold the loss rate grows nearly linearly with energy, $\langle dE_{\rm rad}/dx \rangle = b_w E$, where $b_w \approx 3.4\times10^{-6}~{\rm cm^{-1}}$ (numerically equal to its value in ${\rm g^{-1}cm^2}$, since $\rho_w = 1~{\rm g/cm^3}$) combines muon bremsstrahlung, direct $e^+e^-$ pair production (the most frequent radiative process at these energies) and photonuclear interactions \cite{Groom2001,PDG2020}. The mean radiative deposit per chord is then $\approx 2.0\times10^{-3}\,E$, so $E_\mu \sim 24$--27~TeV in the continuous (mean) energy-loss approximation. Two properties of this approximation deserve comment. First, at multi-TeV energies the radiative losses are strongly stochastic, dominated by rare large energy transfers, so the deposit in an individual tank fluctuates around the mean and its median lies below it; reproducing the observed sustained deposits across $\ge 4$ consecutive tanks with typical rather than rare transfers therefore requires an energy at or above the continuous estimate, which is in this sense conservative. Propagation frameworks implementing full stochastic lepton transport, such as NuLeptonSim, quantify the deviations from the continuous approximation and confirm that neglecting fluctuations overestimates lepton survival \cite{NuLeptonSim2024}. Second, the additional hits produced by these stochastic radiative losses along the track are precisely what enhances the detection efficiency of TeV-scale tracks in HAWC \cite{Albert2022}, consistent with the bright, sustained topologies of the candidates. The charge-space comparison of Section~\ref{sec:chargespace} pushes even higher. The defensible conclusion is $E_\mu \gtrsim 10$~TeV, possibly $\gtrsim 100$~TeV; even the worst-case minimum imposes $E_\mu > 122$--136~GeV, and the conservative $\kappa_v$ reference $E_\mu \gtrsim 200$~GeV.

\subsection{Single catastrophic loss} \label{sec:catastrophic}

The standard objection (a single hard bremsstrahlung of a $\le$100~GeV muon in one tank) fails quantitatively: even if a 100~GeV muon deposited its entire energy in a single tank ($\approx 3300~\pe$), the average over $\ge 4$ pixels would be $\le 850~\pe$, below both candidates, and the remaining pixels would show MIP-compatible charges, contrary to the sustained bright deposits visible in the published displays (Fig.~\ref{fig:displays}).

\begin{figure}[pos=htbp]
\centering

\includegraphics[width=0.95\linewidth]{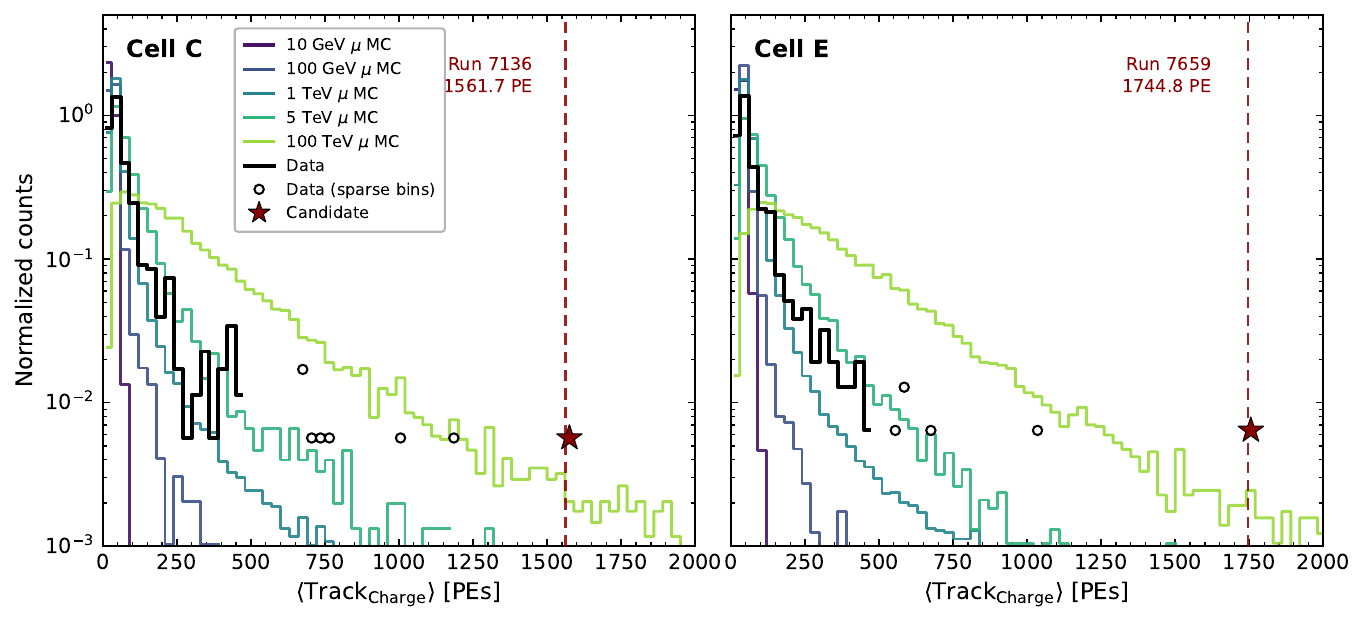} \caption{Distributions of the average charge per pixel for tracks pointing to cells C and E, redrawn from Ref.~\cite{ICRC2019}, compared with the MC expectations for mono-energetic muons. The bulk of the population, consistent with the scattered-muon background of Ref.~\cite{Albert2022}, is separated from the two candidates by more than an order of magnitude in charge; the sparsely populated bins in between are discussed in Section~\ref{sec:threshold}.} \label{fig:charge}
\end{figure}

\subsection{Charge threshold and topological verification} \label{sec:charge} \label{sec:threshold}

The bound of Eq.~(\ref{eq:bound}) defines a charge threshold rather than a pair of events. Any track whose total charge satisfies $Q_{\rm tot} > 100\,\kappa \approx 3.3\times10^{3}~\pe$, that is $\langle q \rangle_{\rm pixel} \gtrsim 830~\pe$ over the minimum four tanks, is inconsistent with the scattered-muon model through the comparison of Eq.~(\ref{eq:exclusion}). Five tracks in the published charge distributions exceed it: in cell C, two at $\approx$1005 and $\approx$1185~$\pe$ besides the candidate at 1561.7~$\pe$, and in cell E one at $\approx$1035~$\pe$ besides the candidate at 1744.8~$\pe$ (the values quoted for the three unanalyzed tracks are bin centers of the published distributions, binned in 30~$\pe$).

The average charge alone cannot promote any of them to a candidate, because it is degenerate with respect to a single anomalous channel: one PMT recording 3930~$\pe$ together with three tanks at the minimum-ionizing level reproduces $\langle q \rangle_{\rm pixel} = 1005~\pe$ exactly, with no high-energy lepton involved. Resolving that degeneracy requires the distribution of charge among tanks, which is public only for the two candidates.

Figure~\ref{fig:displays} provides the comparison, together with a reference from the same instrument: the horizontal muon of Run 7739, with $\langle q \rangle_{\rm pixel} = 55.4~\pe$. Its reconstructed azimuth, $\phi = 254.6 \pm 2.8^\circ$ \cite{ICRC2019}, points far from the volcano, to a direction that the terrain profile leaves essentially unshielded. It is therefore most plausibly a genuine near-horizontal atmospheric muon, reaching the array through atmosphere alone and requiring none of the deflection that defines the scattered-muon background of Ref.~\cite{Albert2022}. As a reference it has the same topology as the candidates, recorded by the same instrument, in a direction with no rock in the way. Thirteen consecutive tanks carry signal along its line and twelve of them have two or more PMT hits, against the published track length of ${\rm TL} = 10$ pixels \cite{ICRC2019}. The tracking criteria require two PMTs above $4~\pe$ per pixel, which the ends of the track do not meet, so ${\rm TL}$ underestimates the physical extent of a track at its extremities. Marker sizes are quoted below as the dimensionless ratio $s = r_{\rm marker} / r_{\rm max}$ of the marker radius to the maximum of the published scale of Fig.~\ref{fig:displays}. That scale is bounded at both ends: every channel below its low end is drawn at the same minimum size, and every channel above its high end is drawn clipped at $s = 1$, a condition we call saturation, so that the display bounds the charge of such a channel only from below. No absolute charge is inferred from $s$: all the statements of this subsection are ordinal or counts of channels and of tanks.

The reference muon reaches $s = 0.92$ in its brightest channel and no more than $s = 0.60$ in any other, and none of its markers saturates. Both candidates saturate, in every case inside the tanks of the track itself: nine markers in Run 7136 and two in Run 7659. In Run 7136 the nine occupy three of the five WCDs of the track, with all four PMTs of one tank and three of the four of the tank adjacent to it saturated, the remaining two in a third tank further along the track; the tenth-largest marker of that event still reaches $s = 0.90$. In Run 7659 the two saturated markers lie in a single WCD of the track. This resolves the degeneracy described in the preceding paragraph for both events: a single anomalous channel accompanied by three tanks at the minimum-ionizing level predicts exactly one marker at the maximum with the rest of the track at the minimum, and neither candidate shows that configuration. The number of saturated markers is not, however, a measure of how many channels are anomalous, because clipping truncates the scale from above; in that respect the verification is stronger for Run 7136, where seven PMTs of two adjacent tanks saturate, than for Run 7659, where it rests on two channels of one tank.

The displays also record signal in tanks adjacent to the track, which the reference muon does not. The array has an almost uniform nearest-neighbor spacing of $7.90$~m, so adjacency is a property of the geometry and not a distance chosen by us; the track axes are defined here by the collinearity of the tank centers carrying signal, satisfied to $\pm 0.4$~m over 48~m in Run 7659 and to $\pm 2.1$~m over 41~m in Run 7136. The reference muon has signal in none of the sixteen tanks adjacent to its thirteen-WCD track, Run 7136 in four of nine and Run 7659 in three of eleven. Among the tanks beyond one spacing from the track the occupancy is the same in the three events, 8.7, 7.9 and 8.0 per cent of the available WCDs, which identifies that population as the uncorrelated vertical background contained in the trigger window. Taking the highest of the three, 8.7 per cent, as a common and conservative estimate of that rate, and treating the adjacent tanks as independent trials, the binomial probability of the observed adjacent occupancy is 0.5 per cent for Run 7136, which has 4 of 9 adjacent tanks occupied against 0.8 expected, and 6 per cent for Run 7659, with 3 of 11 against 1.0 expected. The absence of adjacent signal in the reference muon is not significant: the same rate predicts 1.4 occupied tanks among its 16, and the observed zero has probability 23 per cent. These figures assume the off-track occupancy to be uncorrelated between tanks, as its uniformity across the three events suggests, and are quoted as indicative; the conclusion drawn from them is qualitative: both candidates show adjacent activity at a level the uncorrelated background does not naturally produce, and the reference muon shows none.

The amplitude of the adjacent deposits is perhaps more informative than their frequency. The brightest adjacent-tank marker of Run 7659 reaches $s = 0.98$, with a second at $s = 0.80$ in the same tank, while across the sixty-seven background WCDs of the three displays no marker exceeds $s = 0.70$. A deposit of that size one tank off the track is expected of a radiative loss or of a cascade, whose secondaries are emitted over a range of angles, and is not expected of a minimum-ionizing particle, which deposits in the water column it crosses and in no other. We record the observation as consistent with the former and difficult to reconcile with the latter, without asserting a causal association: establishing that the adjacent deposits belong to the same event would require the relative timing of the hits.

We therefore analyze the two events whose charge distribution can be verified, and make no claim about the remaining three, which are neither confirmed nor excluded. The rate used in Section~\ref{sec:rates} is accordingly two events, which is the conservative choice: counting all five would raise the tension discussed there from $\sim$2.1--2.5$\sigma$ to $\sim$4.5--5.1$\sigma$.

The same threshold applied across the analysis region reinforces the angular argument of Section~\ref{sec:direct}. Because the six cells defined in Section~\ref{sec:inputs} pair the same three azimuth bins in the two elevation rows, the effect of the overburden can be separated from that of the azimuth, and no property of the charge distributions enters the definition of the region.

Within that region the largest average track charge is $1561.7$ and $1744.8~\pe$ in cells C and E, against $\approx$1000, 700, 550 and 450~$\pe$ in cells K, I, J and D, each a single track. The two cells of largest overburden thus contain the two brightest tracks of the six-cell sample, and the candidates exceed the largest charge of the four remaining cells by a factor $1.5$--$1.7$. The track of $\approx$1000~$\pe$ in cell K meets the charge threshold but not the overburden condition, so penetrating atmospheric muons are not excluded there.

The comparison cannot be extended beyond these six cells: Ref.~\cite{ICRC2019} publishes charge distributions for cells I, J, K, C, D and E only, and no statement is made here about the remaining six. One neighbouring cell nevertheless deserves note in advance, because it bounds what such an extension could establish. The ray tracing of Section~\ref{sec:inputs} separates the two rows by a factor of two in shielding: the lower row averages 18.6 to 21.0~$\kmwe$ and the upper row 10.5 to 12.1~$\kmwe$. Ref.~\cite{ICRC2019} states that the cells serve to study the properties of the reconstructed tracks as a function of the average width of volcano that they traverse, and this factor of two is that contrast. Cell H belongs to the upper row, and its average of 10.6~$\kmwe$ is indistinguishable from that of cell K, 10.6~$\kmwe$, the least shielded cell of the analysis region. More to the point, H is the only cell among the seven examined here whose field of view extends past the edge of the volcano: its minimum overburden is zero, and 4.6~per~cent of its solid angle lies below 5~$\kmwe$, whereas every other cell examined has a minimum of at least 7.9~$\kmwe$. A bright track recorded in cell H would therefore require no penetrating capability at all, since part of that cell looks alongside the mountain rather than through it.

Charge above $\approx$1000~$\pe$ is thus recorded in cells K, C and E, and it coincides with a large and uniform overburden only in C and E. Large charge on its own is achieved on both sides of the shielding contrast; the condition met only by cells C and E is large charge and large overburden simultaneously, which is the conjunction the argument of Section~\ref{sec:direct} rests on.

\section{Direct-muon and muon-bundle backgrounds} \label{sec:otherbkg}

\subsection{Direct atmospheric muons} \label{sec:direct}

For the directions of cells C and E the overburden exceeds $18~\kmwe$ ($X = 1.8\times10^{6}~{\rm g/cm^2}$) (several times the vertical overburden of the deepest underground laboratories, albeit restricted to specific horizontal directions) making HAWC one of the best-shielded surface instruments for horizontal cosmic-ray suppression. The minimum surface energy to penetrate it is
\begin{equation}
E_{\rm min} = \frac{a}{b}\left(e^{bX}-1\right) \approx 0.67~{\rm PeV},
\label{eq:emin}
\end{equation}
with $a \approx 2.0\times10^{-3}$~GeV\,cm$^2$/g and $b \approx 4.0\times10^{-6}$~cm$^2$/g \cite{Groom2001}. These energy-loss parameters follow the standard-rock convention ($\rho = 2.65~{\rm g/cm^3}$, $\langle Z/A \rangle = 0.5$) of Menon and Ramana Murthy \cite{MenonMurthy1967}, as adopted in the benchmark tables of the Particle Data Group \cite{PDG2020}. Integrating the IceCube atmospheric-muon spectrum \cite{IceCubeMuons2016} as parametrized in Ref.~\cite{LeonVargas2017} ($\propto E^{-3.73}$) above this threshold gives $\Phi_\mu(>0.67~{\rm PeV}) \approx 1.8\times10^{-3}~{\rm m^{-2}\,sr^{-1}\,yr^{-1}}$, which with $A\Omega = 5.1\times10^{-2}~{\rm m^2\,sr}$ and 181 days yields
\begin{equation}
N_{\rm direct} \approx 4.5\times10^{-5}~{\rm events},
\end{equation}
confirming that direct penetration of atmospheric muons is not a viable background source at these energies.

The robustness of this estimate rests on three considerations, which bound it from both directions. First, the normalization of the near-horizontal muon flux is anchored by direct measurement: the 800-ton MUTRON magnetic spectrometer measured the muon spectrum at $86^\circ$--$90^\circ$ zenith angle up to 20~TeV, with a maximum detectable momentum (MDM) of 22~TeV/$c$ (the highest reported for a horizontal muon spectrometer) obtaining $I_\mu(>1~{\rm TeV}) = (1.70\pm0.10)\times10^{-7}~{\rm cm^{-2}s^{-1}sr^{-1}}$ and a production spectral index $\gamma_\pi = 2.73\pm0.02$ \cite{Matsuno1984}, free of the rock energy-loss uncertainties that affect underground range measurements. The parametrization used above is consistent with this anchor in the measured range. Second, in-rock energy degradation is neglected: the continuous energy-loss approximation of Eq.~(\ref{eq:emin}) ignores the stochastic nature of the radiative processes which dominate above the muon critical energy in standard rock, $E_{\mu c} \approx 693$~GeV \cite{PDG2020}. Simulations with stochastic transport (MMC \cite{ChirkinMMC}) show that these fluctuations reduce the mean muon range, and hence the survival probability at large depths, a conclusion shared by NuLeptonSim \cite{NuLeptonSim2024}; both push the true rate below the analytical bound. Third, and in the opposite direction, the extrapolation to 0.67~PeV (a factor $\sim$30 above the MUTRON measured range) enters the regime where prompt muons from charm decay may dominate the horizontal flux; theoretical predictions for this crossover vary widely, from $\sim$60~TeV to $\sim$2~PeV in the horizontal direction, as evaluated in Ref.~\cite{Matsuno1984} from the models of Elbert, Gaisser and Stanev \cite{Elbert1983} and of Inazawa and Kobayakawa \cite{Inazawa1983}. A prompt-dominated flux is flatter than $E^{-3.73}$, so $N_{\rm direct}$ could increase by a factor of a few to $\sim$10 in the most pessimistic scenario; even then it remains $\lesssim 10^{-3}$ events, while the stochastic suppression of the second consideration acts against it. The conclusion $N_{\rm direct} \ll 1$ is therefore robust from both directions. 

The angular behavior of this background is further constrained by observation. In Super-Kamiokande, the residual muon contamination of the upward-going sample near the horizon was shown to concentrate in the azimuthal directions of minimum overburden of the surrounding mountain profile: clusters of near-horizontal muons appear in the thin-shield directions (``Region (1)'' in their notation), while the thick-overburden directions (``Region (2)'') show no such contamination; the background was quantified and subtracted by exploiting precisely this anticorrelation \cite{SuperK2005}. Cosmic-ray muon backgrounds near the horizon therefore populate preferentially the directions where the natural shield is thinnest. The two HAWC candidates exhibit the opposite behavior: they point to the cells of maximum overburden, $>18~\kmwe$ \cite{ICRC2019}, along the cumulative overburden profile of the Pico de Orizaba \cite{Albert2022}, where any muon background is weakest and where only the neutrino-induced signal, whose rate is essentially independent of the overburden (Section~\ref{sec:formalism}), survives.

\subsection{Collinear muon bundles} \label{sec:bundles}

A bundle of sub-TeV muons could in principle mimic a large sustained charge without violating per-particle energy conservation. Five independent arguments exclude this background:

\emph{(1) Multiplicity requirement.} The per-muon charge relevant here is not the vertical anchor of Table~\ref{tab:inputs}, which corresponds to a $4.5$~m vertical water path, but the charge that the near-horizontal muon population actually deposits in a WCD, $\approx 65$--$84~\pe$ per pixel (Section~\ref{sec:exclusion}). Reproducing $\langle q \rangle_{\rm pixel} = 1560$--$1745~\pe$ therefore requires $N \gtrsim 19$--27 muons per tank sustained over $\ge 4$ consecutive tanks. The published track-geometry cuts ($\ge 4$~m intersection per WCD, linear fit, $\ge 75\%$ of tanks on the line \cite{Albert2022}) confine such a bundle to a cylinder of $\sim$4~m effective radius over $\ge 30$~m, i.e.\ an areal muon density $\rho_\mu \gtrsim 0.4$--2~m$^{-2}$ confined to the near-axis region, the upper end corresponding to a more compact core of $\sim$2~m radius. Throughout this section $\rho_\mu$ denotes a number of particles per unit area, and is not to be confused with the mass density $\rho$. Such densities are extreme for the near-horizontal geometry required here. At large zenith angles the muons reaching the ground are produced far away (the mean production-to-ground path grows from $\sim$10~km for vertical showers to $\sim$300~km for nearly horizontal ones, so that only muons produced above $\sim$100~GeV survive), and the geomagnetic field separates $\mu^{+}$ from $\mu^{-}$ into two lobes on either side of the transverse field component, destroying the compact core \cite{Ave2000}. Both effects dilute the near-axis density by orders of magnitude with respect to a vertical shower of the same primary energy, so that reproducing $\rho_\mu \gtrsim 0.4$--2~m$^{-2}$ over the required track length demands primaries well above the knee. We adopt $E_{\rm CR} \gtrsim 10^{17}$~eV in the rate maximum of item~(5) below, and because that threshold is adopted rather than derived we quantify there how the resulting bound depends on it.

\emph{(2) Geometric blocking.} A shower genuinely pointing to cells C or E cannot exist: its electromagnetic and hadronic components are absorbed within meters of rock. Experimental muon detectors routinely suppress the electromagnetic component of air showers with only $\approx 19$ radiation lengths of soil shielding (e.g., the Tibet underground muon-detector array \cite{Sako2009}); the Orizaba overburden toward cells C and E, $>18~\kmwe$, corresponds to $\approx 7\times10^{4}$ radiation lengths of standard rock. Any muon from such a shower would need $E \gtrsim 0.67$~PeV to penetrate it (Eq.~\ref{eq:emin}), three orders of magnitude above the sub-TeV energy of the individual bundle muons, each of which crosses the tanks as a minimum-ionizing particle. Bundles can only enter by angular migration from unblocked directions.

\emph{(3) Angular migration.} A shower able to deliver such a bundle must arrive from a direction that the mountain does not block, since its electromagnetic and hadronic components are absorbed within meters of rock by item~(2). The terrain profile of Section~\ref{sec:inputs} places the silhouette edge at $13.1^\circ$ and $13.5^\circ$ of elevation at the azimuths of cells C and E, so the migration required in elevation is $9.1^\circ$ and $9.5^\circ$ measured from the upper edge of those cells, and $11.1^\circ$ and $11.5^\circ$ from their centers. Two published figures bracket the elevation resolution. Ref.~\cite{Albert2022} reports $0.7^\circ$ for $\theta_{\rm Rec} < 2^\circ$ within its six high-efficiency azimuth bins; Ref.~\cite{ICRC2019}, whose analysis produced the candidates, quotes an average angular aperture of $3.0^\circ$ between injected and reconstructed muons, and a per-track elevation uncertainty of $\pm 1.9^\circ$ for the reference event of its Table~1. Cell E lies outside the azimuth range of Ref.~\cite{Albert2022}, so the resolution applicable to the candidates is plausibly the poorer of the two, and we adopt the $3.0^\circ$ value: the required migration is then a $3.0$--$3.8\sigma$ displacement, against the $13$--$16\sigma$ that $0.7^\circ$ would give. The two published resolutions differ by a factor of four, not by orders of magnitude, and the migration is required independently of each of the two candidates. The angular error of a bundle reconstructed as a track is moreover smaller, not larger, than that of a single muon: the $\sim$20 near-parallel particles crossing each tank deposit twenty times the light of a single crossing, so the per-tank time and charge centroids that enter the linear fit are correspondingly better determined.

A second constraint is independent of the reconstruction altogether. The array sits on a flat surface \cite{Albert2022}, so a straight track depositing light in the water of tanks whose centers are separated by a horizontal distance $L$ cannot change height by more than the water depth $d$ between its crossings of the first and last tank; since each crossing can lie up to one tank radius inside the span, the horizontal separation of the crossings is at least $L - 2r$ with $r = 3.65$~m, which caps the elevation at $\theta \le \arctan[d/(L - 2r)]$. The published values for $d$ span $4.0$~m of water above the PMTs \cite{Smith2015}, $4.5$~m \cite{HAWCDetector2023,ICRC2019} and $4.8$~m implied by a volume of $2\times10^{5}$~liters over the $41.9~{\rm m^2}$ tank area \cite{Albert2022,Smith2015}; we adopt the largest, which gives the weakest bound. The tank-center spans of the two candidates are 41 and 48~m, giving $\theta \le 8.1^\circ$ and $6.7^\circ$ respectively, a factor of $1.6$--$2$ below the silhouette edge; these are ceilings on any track of that length, not estimates of the candidate elevations, which are reconstructed near $1^\circ$. The spans are measured from the published displays, and a longer one would only tighten the bound. At those maximum elevations the overburden is still $8$--$9~\kmwe$. The platform is graded to a residual slope of $\approx$1.0\% \cite{HAWCDetector2023}; adding its full contribution $0.01\,L$ to the height budget relaxes the caps only to $8.8^\circ$ and $7.4^\circ$, a factor of $1.5$--$1.8$ below the silhouette edge, with $7$--$8~\kmwe$ of overburden remaining. A shower arriving from an unblocked direction therefore cannot produce a track of the observed length at all, however well or badly its elevation is reconstructed. The same relation applied to the reference muon of Run 7739, whose track spans 102~m, gives $\theta \le 2.9^\circ$, consistent with the $1.0 \pm 1.9^\circ$ reported for it in Table~1 of Ref.~\cite{ICRC2019}.

\emph{(4) Topology vetoes.} A dense muon core never arrives alone: its accompanying halo activates WCDs off the track line and violates the $\le 100$ active PMTs condition and the isolation criteria defined by the Hit Activity and Multiple Hit Activity variables (${\rm HA} < 5.65$, ${\rm MHA} < 1.5$), documented to reject 99.93\% of track candidates with $<1\%$ false-positive rate \cite{Albert2022}. The selection that produced the candidates used the same two variables under their earlier names, with ${\rm LC} < 6$ and ${\rm HC} \le 1.5$ \cite{ICRC2019}; the argument holds under either set of values. No plausible geometry allows a $\rho_\mu \gtrsim 0.4~{\rm m^{-2}}$ core to cross the array in a straight line while its halo, arriving in the same 1.5~$\mu$s trigger window, remains below these thresholds.

\emph{(5) Raw rate maximum.} Even ignoring all the above, the flux of capable primaries bounds the background. Let $E_{\rm CR}$ denote the minimum primary energy assumed capable of producing the required bundle. With $E_{\rm CR} = 10^{17}$~eV, $\Phi_{\rm CR}(>E_{\rm CR}) \approx 1.6\times10^{-10}~{\rm m^{-2}s^{-1}sr^{-1}}$, obtained by direct integration of the spectrum published near the second knee of the cosmic-ray spectrum \cite{TALE2018}, the lateral area of the array seen horizontally ($\approx 140~{\rm m} \times 5~{\rm m} \approx 700~{\rm m^2}$), the combined solid angle of the two cells ($\approx 1.5\times10^{-2}$~sr) and $T = 181$~days give $N_{\rm raw} \approx 2.6\times10^{-2}$ events. Lowering $E_{\rm CR}$ raises $N_{\rm raw}$ in proportion to the integral flux: direct integration of the published spectrum gives $\Phi_{\rm CR}(>E_{\rm CR})$ larger by factors of 4.1 and 11.1 for $E_{\rm CR} = 5\times10^{16}$ and $3\times10^{16}$~eV, matching the scalings of 4.0 and 11.1 expected for an integral index of $2.0$ near the second knee, so that $N_{\rm raw}$ becomes $0.10$ and $0.29$ events. The raw maximum therefore stays below $0.3$ events in the 181-day sample for any $E_{\rm CR}$ above $3\times10^{16}$~eV, and none of the conclusions below depends on which of the three values is adopted. Two of the requirements listed above can be folded in explicitly. First, the core-impact condition of item~(1): the axis must cross within $\sim$4~m of the line joining $\ge 4$ WCDs, so that only $\approx 50~{\rm m^2}$ of the $700~{\rm m^2}$ lateral cross-section is useful, $f_{\rm core} \approx 7\times10^{-2}$. This is a deliberately generous estimate: the strict chord condition, which for a $4$~m intersection in a tank of radius $3.65$~m admits impact parameters up to $3.05$~m, confines the axis to a $6.1$~m strip over the $4.5$~m water depth, or $\approx 28~{\rm m^2}$, $f_{\rm core} \approx 4\times10^{-2}$. Second, the published rejection power of the track-isolation cuts of item~(4), $f_{\rm veto} = 1 - 0.9993 = 7\times10^{-4}$. This efficiency was measured on the track-candidate population actually recorded by HAWC, whose multiplicity lies far below $N \approx 19$--27, so transferring it to a dense bundle is an extrapolation, but a conservative one, since a $\rho_\mu \gtrsim 0.4~{\rm m^{-2}}$ core is harder, not easier, to hide from the HA/MHA variables than a sparse one. The product is
\begin{equation}
N_{\rm bundle} \lesssim N_{\rm raw}\,f_{\rm core}\,f_{\rm veto}
\approx 1.3\times10^{-6}~{\rm events}.
\label{eq:nbundle}
\end{equation}
The angular migration of item~(3) is a further suppression that we deliberately leave out of Eq.~(\ref{eq:nbundle}): a $3.0$--$3.8\sigma$ elevation displacement would contribute a Gaussian factor of $1.3\times10^{-3}$ to $7\times10^{-5}$ per event, and $\sim10^{-6}$ for the two required jointly. The non-Gaussian tails of the angular reconstruction are not characterised in the published material, however, and the bound above does not rely on them. It is worth noting that the exclusion does not rest on items~(1) and~(5): the geometric blocking of item~(2), which requires $E_\mu \gtrsim 0.67$~PeV of any muon reaching cells C or E, the $3.0$--$3.8\sigma$ elevation migration of item~(3) and the topology vetoes of item~(4) are independent of the bundle multiplicity and are unaffected by the value adopted for it.

\emph{Empirical closure: absence of a continuum.} A bundle background has a steeply falling, continuous multiplicity spectrum: events at the $N \approx 20$ that the candidate charges correspond to would be accompanied by far more numerous events at $N = 3$, 5 and 10, populating the 200--850~$\pe$ range. The published charge distributions for cells C, D and E (Figs.~6--8 of Ref.~\cite{ICRC2019}) show the opposite morphology: a background population concentrated at $\langle q \rangle \approx 65$--$84~\pe$, a steeply falling intermediate region in which fewer than 2\% of the tracks of either cell lie above 500~$\pe$, and a gap between 765 and 1005~$\pe$ in cell C and between 675 and 1035~$\pe$ in cell E, above which only the isolated tracks of Section~\ref{sec:threshold} remain (Fig.~\ref{fig:charge}). This is the signature of two distinct physical populations, not of the tail of any multiplicity continuum, and it retroactively reinforces the scattered-muon exclusion as well: the candidates are not the tail of any observed population.

\subsection{Instrumental effects} \label{sec:instrumental}

A class of backgrounds not covered by the preceding sections is instrumental: spurious light from electrical discharges or light leaks. Three properties of the candidates disfavor this origin. First, the topology: an instrumental source produces light in a single channel or a single tank, whereas both candidates deposit large, sustained charges in $\ge 4$--5 WCDs whose centers are collinear to $\pm 0.4$--2.1~m over 41--48~m (Section~\ref{sec:threshold}), with the arrival-time gradient encoded in the published displays progressing along the track direction. Second, the selection: both events survived the HA and MHA topological cuts,
which reject 99.93\% of triggers with a false-positive rate below 1\%
\cite{Albert2022} and were applied identically to the whole population from which the published charge distributions are drawn; an instrumental artifact able to imitate collinear multi-tank tracks would moreover populate all azimuths, with no reason to prefer the two cells of maximum overburden. Third, the same instrument, within the same selection, recorded the reference muon of Run 7739 with the same topology and no saturated markers, showing that the tracking selection does not by itself manufacture bright tracks. A definitive exclusion of instrumental pathologies requires non-public monitoring information available only to the collaboration, but nothing in the public record points in that direction.

\section{Expected neutrino-induced signal} \label{sec:signal}

\subsection{Formalism and identifiability window} \label{sec:formalism}

For the $\nu_\mu \to \mu$ channel, let $E_\mu$ denote the energy with which the muon emerges from the rock, and $E_{\rm th}$ the threshold above which the resulting event is identifiable ($E_{\rm th} = 5$~TeV in Table~\ref{tab:pnu}). The probability that a neutrino of energy $E_\nu$ produces a muon with $E_\mu > E_{\rm th}$ is
\begin{equation}
P(E_\nu; E_{\rm th}) = N_A\,\rho\,\sigma_{\rm CC}(E_\nu)\,
\min\!\left[L,\,L_{\rm eff}(E_\nu; E_{\rm th})\right],
\qquad
L_{\rm eff} = \frac{1}{b\rho}\,
\ln\!\left[\frac{E_{\mu 0}+\epsilon}{E_{\rm th}+\epsilon}\right],
\label{eq:prob}
\end{equation}
where $L$ is the thickness of rock available along the line of sight, $N_A$ is Avogadro's number, and $\rho$ together with the energy-loss coefficients $a$ and $b$ are the standard-rock values introduced with Eq.~(\ref{eq:emin}). Here $E_{\mu 0} = (1-\langle y \rangle)E_\nu \approx 0.65\,E_\nu$ is the muon energy at the interaction vertex, $\langle y \rangle$ being the mean inelasticity of the charged-current interaction, $\epsilon = a/b = 500$~GeV,\footnote{The parameter $\epsilon = a/b$ is the same ratio that defines the muon critical energy. With the constant coefficients adopted here it takes the value 500~GeV, whereas the self-consistent crossing of the energy-dependent $a(E)$ and $b(E)E$ quoted in Section~\ref{sec:direct} gives $E_{\mu c} \approx 693$~GeV. The difference enters $L_{\rm eff}$ only inside a logarithm and shifts it by a few per cent over the energy range of interest.} $1/(b\rho) = 0.94$~km, and $\sigma_{\rm CC} = 5.53\times10^{-36}\,(E_\nu/{\rm GeV})^{0.363}~{\rm cm^2}$ \cite{Gandhi1998,LeonVargas2017}. The power-law form (of $\sigma_{\rm CC}$) is reported in Ref.~\cite{Gandhi1998} for $10^{7} \le E_\nu \le 10^{12}$~GeV; below that range it overestimates the tabulated CTEQ4--DIS cross sections of the same reference, by a factor of 3.4 at 10~TeV, 1.8 at 100~TeV and 1.3 at 1~PeV, converging to within 10\% at 10~PeV. It is retained here for continuity with the rate estimate of Ref.~\cite{LeonVargas2017}, and the effect of the tabulated values is quantified below.\footnote{Recomputing Table~\ref{tab:pnu} with the tabulated cross sections would reduce the muon-channel identifiable rate and the intensity $I(E_\mu > 5~\mathrm{TeV})$ by a factor of $\approx$2.5--3, while leaving the tau channel, which operates in the multi-PeV regime where the parametrization is accurate, essentially unchanged. The lower end of $\lambda_{\rm tot}$ is unaffected, since the muon channel is negligible there, and the upper end is an allowance rather than a value derived from this chain, so it does not scale with the cross sections either; the range of Section~\ref{sec:rates} therefore stands. Correcting the cross sections and the flux together, rather than one at a time, changes $I(E_\mu > 5~\mathrm{TeV})$ by only $\approx$15\%, since the two approximations bias the result in opposite directions. The direction of every conclusion of this work is unaffected.} The factor $L_{\rm eff}$ is the thickness of rock within which an interaction still delivers a muon above threshold, and follows from integrating the continuous-loss approximation $-dE/dX = a + bE$ \cite{LipariStanev1991}. It is a property of the muon, not of the target: only interactions occurring within the last $L_{\rm eff}$ of rock before the exit surface deliver a muon above threshold, so a target thinner than $L_{\rm eff}$ contributes over its full thickness $L$ instead. The $18~\kmwe$ of Section~\ref{sec:direct} correspond to $L = 6.8$~km of standard rock, which exceeds $L_{\rm eff}$ for every $E_\nu$ below $\sim$11~PeV, so the thick-target limit $P = N_A\,\rho\,\sigma_{\rm CC}\,L_{\rm eff}$ is the branch that applies here. At and above $\sim$10~PeV, where the power-law form of $\sigma_{\rm CC}$ is declared valid, it remains consistent with the modern BGR18 benchmark, which incorporates higher-order QCD corrections and small-$x$ resummation \cite{BGR18}; at 10--100~TeV, BGR18 agrees with the tabulated cross sections of Ref.~\cite{Gandhi1998} rather than with the power-law extrapolation. The focus on the $\nu_\mu \to \mu$ channel is itself supported by stochastic propagation studies: below $\sim$100~PeV the Earth-emergence probability of muons exceeds that of taus, for both $\nu_\mu$ and $\nu_\tau$ primaries \cite{NuLeptonSim2024}. The effective target is not the mountain but the muon range, which grows only logarithmically with energy. Setting $E_{\rm th} \to 0$ in $L_{\rm eff}$ recovers the full range of a muon of energy $E_\mu$, $R(E_\mu) = (1/b\rho)\ln(1+E_\mu/\epsilon)$: approximately 1~km at 1~TeV, 3~km at 10~TeV, and 7~km at 1~PeV in standard rock, with detailed stochastic Monte Carlo (MMC) confirming this scale while predicting somewhat shorter effective ranges due to fluctuations \cite{ChirkinMMC}. The signal is therefore essentially independent of the overburden once the latter exceeds this range ($\sim$1--3~km for the relevant energies), while every background falls with overburden. The discrimination of the signal is optimal at the base of the volcano, where both candidates appeared.

An event is identifiable only if its charge stands out from the scattered-muon population subsequently characterized in Ref.~\cite{Albert2022}. Deposits of the magnitude recorded for the candidates require the radiative losses of a multi-TeV muon, and comparison with the mono-energetic MC samples published in Ref.~\cite{ICRC2019} places that scale at $E_\mu \gtrsim 5$--10~TeV. We adopt $E_{\rm th} = 5$~TeV throughout, the lower end of that range. We choose this value since it lies well below the energies that the candidate charges themselves imply (Section~\ref{sec:exclusion}), so that a lower threshold maximizes the predicted rate against which the excess is measured. It also places the predicted median below the charge-inferred band of Section~\ref{sec:exclusion} rather than inside it. The scattered-muon intensity peaks at $\approx 4$~GeV and is confined below $\sim$100~GeV \cite{Albert2022}, so neutrino-induced events dominate the identifiable region by construction. We adopt the analytic approximation of Table~\ref{tab:inputs} for the conventional horizontal atmospheric $\nu_\mu + \bar\nu_\mu$ flux. Its normalization agrees to within 20\% at 1~TeV with the standard pion--kaon parametrization \cite{Volkova1980,GaisserHonda2002} evaluated at the horizon, and its slope matches the index measured by IceCube over the 10--100~TeV range that dominates the identifiable signal \cite{IC40}. That measurement is a zenith average over $97^\circ$--$180^\circ$ and lies a factor 1.3--1.7 below the present approximation in that range, as expected for a sample from which the near-horizontal band was excluded. Convolving Eq.~(\ref{eq:prob}) with this flux gives a distribution of identifiable events with median $E_\nu \approx 16$~TeV, a central 68\% range of 10--32~TeV, and median $E_\mu \approx 7$~TeV (Table~\ref{tab:pnu}).

\begin{table}[pos=htbp]
\centering \caption{Conversion probability and contribution to identifiable events ($E_{\rm th} = 5$~TeV) for the $\nu_\mu \to \mu$ channel. The last column is $\Phi_\nu\,P\,E_\nu$, the contribution per unit $\ln E_\nu$ evaluated at the tabulated energies; integrating it over $\ln E_\nu$ reproduces $I(E_\mu > 5~{\rm TeV})$ of Section~\ref{sec:rates}.} \label{tab:pnu}
\begin{tabular}{c c c c c}
\hline
$E_\nu$ & $\sigma_{\rm CC}$ [cm$^2$] & $L_{\rm eff}$ [km] & $P(E_\nu; >5~{\rm
TeV})$ & ${\rm d}I/{\rm d}\ln E_\nu$ [cm$^{-2}$s$^{-1}$sr$^{-1}$] \\
\hline
10 TeV  & $1.6\times10^{-34}$ & 0.23 & $5.6\times10^{-6}$ & $1.1\times10^{-15}$ \\
30 TeV  & $2.3\times10^{-34}$ & 1.22 & $4.5\times10^{-5}$ & $4.7\times10^{-16}$ \\
100 TeV & $3.6\times10^{-34}$ & 2.33 & $1.3\times10^{-4}$ & $5.4\times10^{-17}$ \\
\hline
\end{tabular}
\end{table}

Table~\ref{tab:pnu} gives the intensity of identifiable events. The neutrino hypothesis does not predict identifiable events uniformly across energy: it concentrates them between a few and a few tens of TeV of emerging lepton energy, with median $E_\mu \approx 7$~TeV. Below $\sim$5~TeV neutrino-induced events are not separable from the scattered-muon population: the published MC mean charge is $109~\pe$ at 5~TeV and $63~\pe$ at 1~TeV, against the $\approx 65$--$84~\pe$ of the scattered-muon population, so the two overlap. Ionization dominates the energy loss below $\sim$1~TeV and the deposited charge is nearly independent of energy there; only above $\sim$10~TeV does the radiative growth of the deposit separate the two. Far above 100~TeV the rates collapse, driven by the steepening of the atmospheric neutrino spectrum ($\propto E^{-3.7}$). The mountain itself remains essentially transparent to neutrinos at these energies: the charged-current conversion probability over the full chord is only $\sim 10^{-3}$ even along the longest one, so the collapse is spectral, not absorptive; the same transparency is what makes the target efficiency of Eq.~(\ref{eq:prob}) linear in $\sigma_{\rm CC}$.

The energy inferred from the candidate charges is a lower bound, and it is worth being explicit why. The chain of Section~\ref{sec:exclusion} converts charge into deposited energy through a calibration constant, and Section~\ref{sec:keff} measures that constant to vary by a factor $\approx 20$ across the published brightness range, from $\kappa_v = 33.3~\pe/{\rm GeV}$ in the minimum-ionizing regime to $\kappa_{\rm eff} \approx 1.5$--$1.7~\pe/{\rm GeV}$ at the brightest published MC point. Adopting the largest of those values minimizes the inferred energy and gives $E_{\rm dep} \gtrsim 190$ and $210~{\rm GeV}$, that is $\approx 47$--$52$~GeV in each of four tanks, which a muon sustains radiatively only if $E_\mu \gtrsim 24$--$27$~TeV. Adopting instead the calibration of the brightness regime that the candidates actually populate, the same chain returns $E_{\rm dep} \approx 4$--$5$~TeV and $E_\mu \approx 5\times10^{2}$~TeV. Both endpoints use published calibration values and neither extrapolates beyond the simulated range. We therefore quote the minimum alone: the candidates require an emerging lepton of more than about 20~TeV, and the published material does not bound the value from above. The interval between the two readings is a systematic of the calibration regime, measured in Section~\ref{sec:keff}, and not an uncertainty on a measurement; for that reason no single energy is quoted anywhere in this work, and no significance is attached to the fact that the charge-inferred minimum of $\approx 20$~TeV and the median $E_\mu \approx 7$~TeV predicted in Table~\ref{tab:pnu} fall within a factor of three of each other. What the two have in common is only that they are compatible, which is a necessary condition for the neutrino interpretation and not evidence for it. The exclusion of the alternatives rests on the event-by-event arguments of Sections~\ref{sec:exclusion} and~\ref{sec:otherbkg}, not on this compatibility.

\subsection{The tau channel and the 2017 prediction} \label{sec:tau}

Ref.~\cite{LeonVargas2017} predicted that $\tau$-induced signals (the collimated decay products impacting essentially a single row of tanks after $\sim$2~km of propagation from the volcano edge) would deposit thousands to tens of thousands of PE. The lower portion of that window falls within the dynamic range of the electronics, which extends up to thousands of PE; for larger deposits, the same reference anticipated that the collected charge would still provide at least a lower bound on the lepton energy. The candidates, at 1562 and 1745~$\pe$ per pixel, fall at the lower edge of that predicted window, which is also where a steep spectrum concentrates the observable events. 

\begin{figure}[pos=htbp]
\centering

\includegraphics[width=0.95\linewidth]{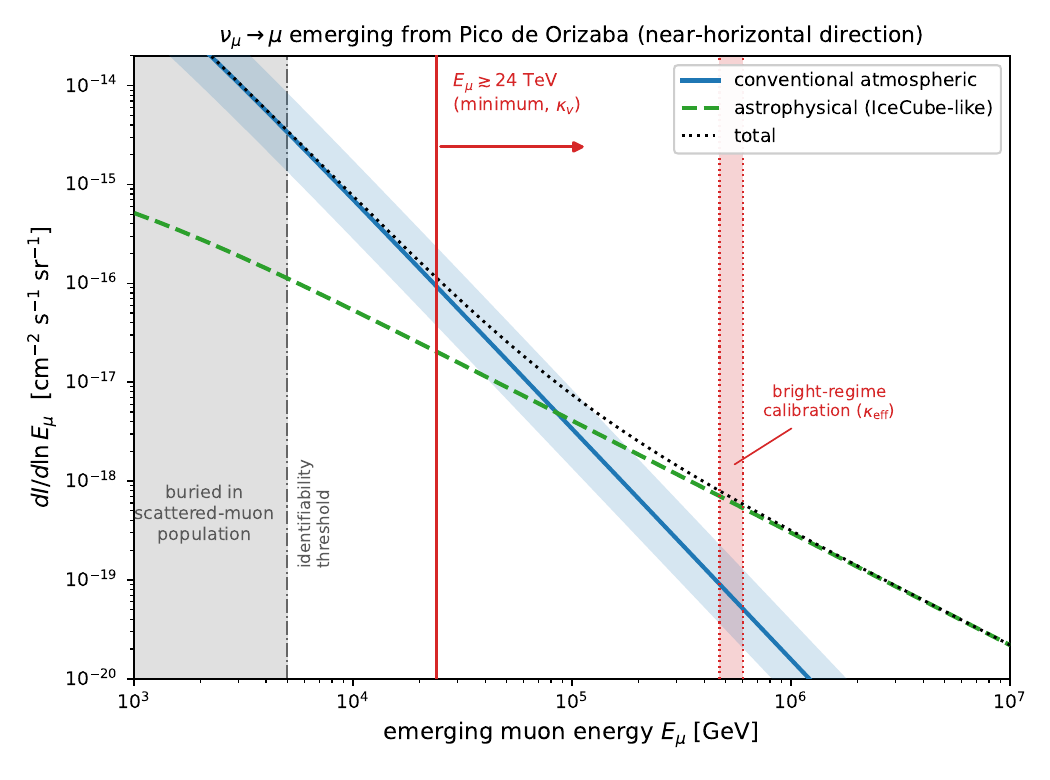} \caption{Expected intensity of identifiable neutrino-induced muons per unit $\ln E_\mu$, emerging from the Pico de Orizaba along the near-horizontal direction, for the conventional atmospheric flux (band: factor 2--3 normalization uncertainty) and the astrophysical extrapolation. The abscissa is the energy $E_\mu$ of the emerging muon, not the neutrino energy of Table~\ref{tab:pnu}. The two red markers show what the candidate charges imply under the two admissible calibrations of Section~\ref{sec:keff}: the solid line with the arrow is the minimum obtained with the most conservative value, $\kappa_v = 33.3~\pe/{\rm GeV}$, and the dotted band is the result of the same chain with the calibration of the brightness regime the candidates populate, $\kappa_{\rm eff} = 1.5$--$1.7~\pe/{\rm GeV}$. The separation between them is the calibration systematic measured in Section~\ref{sec:keff} and not an uncertainty band on a measured energy; only the minimum is quoted in the text. The ordinate is a differential in $\ln E_\mu$, so the intensity above a threshold $E_0$ is $f(E_0)/\gamma$, the value of the curve at $E_0$ divided by its logarithmic slope, and not the area perceived on the logarithmic axes: at $E_0 = 5$~TeV the conventional atmospheric curve reads $f(E_0) = 3.4\times10^{-15}$ and $\gamma = 2.3$, giving the $I(E_\mu > 5~{\rm TeV}) \approx 1.5\times10^{-15}~{\rm cm^{-2}s^{-1}sr^{-1}}$ of Section~\ref{sec:rates}.} \label{fig:spectrum}
\end{figure}

\subsection{Rates} \label{sec:rates}

For the $\mu$ channel, the intensity of identifiable events is $I(E_\mu>5~{\rm TeV}) \approx 1.5\times10^{-15}~{\rm cm^{-2}s^{-1}sr^{-1}}$ for the conventional atmospheric flux (Fig.~\ref{fig:spectrum}). The astrophysical component of Table~\ref{tab:inputs}, whose spectral index is intermediate between, and whose normalization is slightly below, the two IceCube diffuse measurements \cite{IceCubeTG2022,IceCubeST2024}, in both cases within their quoted uncertainties, adding a further $0.1\times10^{-15}$: with the low-energy acceptance this gives $\lambda_\mu \sim 1.2\times10^{-5}$ events in 181 days, rising to $\sim 1.2\times10^{-3}$ with the high-energy acceptance growth of ``at least a couple of orders of magnitude'' documented in Ref.~\cite{Albert2022} (an increase driven by the additional secondary particles from radiative energy losses, which enhance the trigger probability of the modular array). Stretching in addition the factor 2--3 flux normalization gives $\lambda_\mu \lesssim 4\times10^{-3}$. We nevertheless retain the more generous allowance $\lambda_\mu \lesssim 0.1$, a deliberate choice rather than a derived value: it grants the acceptance growth a factor of 25 beyond the traceable stretch, and keeps the $2.1\sigma$ end of the tension range conservative. An independent bound places it well inside what the data permit: LVD measured the neutrino-induced muon intensity above $E_\mu = 1$~GeV at $\theta \approx 90^\circ$ to be $(8.3 \pm 2.6)\times10^{-13}~{\rm cm^{-2}s^{-1}sr^{-1}}$, and showed it to be independent of slant depth beyond $14{,}000~{\rm hg/cm^2}$ of standard rock, so that the same value applies to any detector viewing at that angle through more than $14~\kmwe$ \cite{LVD1995}. Cells C and E, at 18.6 and $21.0~\kmwe$, are in that regime, and with the stretched high-energy aperture the measured intensity caps the expectation of neutrino-induced muons above 1~GeV at $\lesssim 0.7$ in 181 days. For the $\tau$ channel, Table~7 of Ref.~\cite{LeonVargas2017} (trigger $\ge 4$ WCDs, IceCube extrapolation, both energy bins) gives $\approx 0.22$ events/yr, i.e.\ a Poisson expectation $\lambda_\tau = 0.11$ in 181 days (itself an optimistic estimate: the flux parameters of that extrapolation were shifted to the highest values allowed by the IceCube statistical uncertainties). The comparison between the two channels is one of detectability rather than of emergence: below $\sim$100~PeV more muons than taus emerge from the rock \cite{NuLeptonSim2024}, but an emerging $\tau$ that decays in or near the array converts a large fraction of its energy into a compact shower with high trigger efficiency, whereas a muon deposits only $\sim 2\times10^{-3}$ of its energy per crossed tank (Section~\ref{sec:formalism}); the radiative losses of the $\tau$, suppressed by $(m_\tau/m_\mu)^2$ relative to the muon, also allow it to reach the array with most of its energy intact.

The total expectation is the sum of the two channels, $\lambda_{\rm tot} = \lambda_\mu + \lambda_\tau$, and is quoted as a range because $\lambda_\mu$ is not a single number: the acceptance for identifiable events is known only through the statement of Ref.~\cite{Albert2022} that it grows with energy by at least a couple of orders of magnitude, and the flux normalization carries its own uncertainty. The lower end, $\lambda_{\rm tot} \approx 0.11$, takes $\lambda_\mu$ at its nominal value, where it is negligible against $\lambda_\tau$; the upper end, $\lambda_{\rm tot} \approx 0.2$, takes $\lambda_\mu$ at the largest value the published material permits. The range therefore spans model assumptions rather than a statistical uncertainty, and $\lambda_\tau$ is held throughout at the optimistic value quoted above, a choice that inflates $\lambda_{\rm tot}$ and so understates the tension.

With $\lambda_{\rm tot} \approx 0.11$--0.2, the Poisson probability of observing $\ge 2$ events is 0.6--1.7\%. The honest reading is that the observed rate exceeds the neutrino hypothesis by a factor $\sim$10--20 in its central estimate, a $\sim$2.1--2.5$\sigma$ tension with a natural physical outlet already written in Ref.~\cite{LeonVargas2017}: a single event with $E_\nu > 10$~PeV would evidence a flux beyond current extrapolations \cite{Kistler2016}. Such a flux is a central topic of the 2022 Snowmass reports, which emphasize that current bounds from IceCube and Auger leave a vast discovery space in the PeV--EeV range, where new astrophysical components or exotic contributions (such as super-heavy dark-matter decay) could emerge \cite{SnowmassHE,SnowmassNF10}. The tension is $2.5\sigma$ for $\lambda_{\rm tot} = 0.11$ and $2.1\sigma$ for $\lambda_{\rm tot} = 0.2$; reaching $1\sigma$ would require $\lambda_{\rm tot} = 0.71$, more than three times the upper end. It is therefore not an artifact of the acceptance or of the flux assumptions, although with two events it remains a mild one that neither refutes nor confirms the hypothesis. In any case, the background exclusion of Sections~\ref{sec:exclusion}--\ref{sec:otherbkg} is independent of this discussion.

\section{Discussion} \label{sec:discussion}

The inference chain assembled from the HAWC published material alone is: (i) the three conceivable backgrounds assign zero or $\lesssim 10^{-3}$ probability to the candidates; (ii) the neutrino hypothesis predicts identifiable events exclusively in the charge window $10^{3}$--$10^{4}~\pe$ and in the highest-overburden cells, both observed; (iii) the charges require an emerging lepton above about 20~TeV, a minimum that lies above the whole scattered-muon population and inside the range where the calculated emerging spectrum places identifiable events, while the upper end of the energy is left open by the factor of 20 spanned by the published calibration (Section~\ref{sec:formalism}); (iv) the rate shows a $\sim$2.1--2.5$\sigma$ excess, which under the conventional atmospheric flux requires the acceptance and the flux to sit at the optimistic end of their ranges, and which would instead point to a multi-PeV component if the flux is larger than the conventional extrapolation. We note only in passing that a $\tau$-induced shower would also fall in the observed charge window, as anticipated in Ref.~\cite{LeonVargas2017}; the present data cannot discriminate the flavor and nothing in this work rests on it. The multi-PeV regime is also the target of upcoming instruments such as PUEO and TAMBO, designed to resolve whether such an excess corresponds to a new population of cosmic accelerators or to cosmogenic (GZK) neutrinos \cite{PUEO2021,TAMBO2026}.

The contemporary context shows how demanding the near-horizontal regime is. The ultra-high-energy event KM3-230213A, a through-going near-horizontal muon track reported by KM3NeT \cite{KM3NeT2025}, is in quantified tension with the IceCube non-observation over a much larger exposure, a Bayes factor of 18, corresponding to roughly $2.8\sigma$, when both experiments are confronted with the same diffuse flux \cite{Palmisano2026}. The near-horizon, Earth-skimming candidates reported by ANITA-IV provide a further illustration, albeit a physically distinct one: analyzed jointly with KM3-230213A and the IceCube null result, their counts cannot be reconciled with a common Standard-Model neutrino flux at the $5.9$--$7.9\sigma$ level \cite{Chattopadhyay2026}. That anomaly operates at far higher energies (EeV rather than the TeV--PeV regime relevant here) and stands in the opposite relation to IceCube: the flux required for the ANITA-IV events would overproduce IceCube events, whereas the candidates discussed here lie at energies where the required flux is the one IceCube measures.

The historical antecedent is the Kolar Gold Fields program: there too, a small number of anomalous track events, recorded behind a large shielding mass, resisted explanation by the identified backgrounds. The Kolar anomaly was topological, multi-track events with large opening angles and vertices in the air or in thin detector material \cite{KGF1975}, whereas the present one concerns the rate. Anomalous track events unexplained by conventional expectations are thus a recurring feature of the field, from the Kolar events to the present, arising from causes specific to each experiment rather than a shared origin. The use of the horizontal direction as a natural electromagnetic filter is itself well established: at large zenith angles the electromagnetic shower component is absorbed by the large atmospheric slant depth, leaving muon-dominated fronts, the basis of the inclined-shower analyses of surface arrays \cite{Auger2009}.

Limitations. The sample consists of two events. The fidelity of the HAWC MC at the extreme of the charge scale (saturation modeling above $\sim$1500~$\pe$) is the residual systematic, mitigated by the published dynamic range and by the conservative direction of any saturation imperfection. HAWC's front-end electronics measure charge via the time-over-threshold (ToT) method, calibrated in situ from $\sim$0.1 to more than 1000~$\pe$, in which large pulses are encoded through the width of the discriminated signal (four-edge events), so that amplitude clipping of the analog pulse does not by itself exhaust the charge estimate \cite{HAWCDetector2023}. The calibration chain is documented to cover the PMTs from fractions to thousands of PE, and the published charge-rate distribution of a single PMT in triggered data extends to $\sim 10^{4}~\pe$ \cite{HAWCCalib2015}, so per-pixel charges at the scale of the candidates lie well within the routinely calibrated range. The published event displays \cite{ICRC2019} already show qualitatively uniform, large deposits across the tanks along each track, consistent with the sustained-radiative-loss hypothesis; the exact per-tank charge values are not public, and their eventual publication would allow a quantitative per-tank test, strengthening (not enabling) the present conclusions.

Outlook. The statistical significance of the present result is limited by the short exposure analyzed here, and the most immediate way to strengthen it is a modest increase in livetime. Because the identifiable-event search is essentially background-free in the shielded near-horizon window, the expected signal scales approximately linearly with exposure rather than as its square root, so even a factor of a few is discriminating: a genuine excess at the present rate would grow toward evidence level, whereas an upward statistical fluctuation would regress toward the conventional expectation. Three elements indicate that such an extension is practical. First, a data sample spanning roughly three times the present livetime, reconstructed within the same selection and reconstruction framework and with stricter detector-stability requirements, has already been assembled and analyzed for a related horizontal-muon study \cite{Castellon2025}, demonstrating that the extended-exposure analysis is operationally and computationally feasible. Second, that study finds the horizontal-muon rate to be stable over the corresponding 1.5-year period (constant-rate fits with $\chi^2/\mathrm{NDF}$ between 0.85 and 1.10), which constrains the detector-stability systematics most relevant for a longer-term rate analysis. Third, an independent identification method for horizontal tracks of arbitrary azimuth, based on a convolutional neural network that takes the event displays themselves as input, has been developed within the collaboration \cite{AngelesCamacho2021}; it shares the upstream calibration and hit reconstruction with the track finder that produced the candidates, but replaces its explicit topological criteria with a learned classifier, so a re-analysis using both would test the selection against a method with different failure modes. Extending the identifiable-neutrino search to this larger exposure is a decision for the HAWC collaboration; the present work only establishes it as the natural next step. Analysing the complete multi-year HAWC dataset would provide the ultimate test, but is considerably more demanding in computing resources.

\section{Conclusions} \label{sec:conclusions}

We have shown, using exclusively published information, that the two track-like events reported in Ref.~\cite{ICRC2019} are inconsistent with all identified backgrounds. The scattered-muon background is excluded kinematically by energy conservation. Direct atmospheric muons contribute $4.5\times10^{-5}$ expected events, an estimate anchored by the MUTRON horizontal-spectrometer measurement and robust from both directions: stochastic in-rock energy losses push it further down, while even the most pessimistic prompt-charm flux raises it only to $\lesssim 10^{-3}$. Collinear muon bundles are excluded by five independent arguments and by the absence of any multiplicity continuum in the published charge distributions. The candidates moreover exhibit the angular behavior opposite to that of every known muonic contamination: whereas residual near-horizontal backgrounds concentrate in thin-overburden directions, as demonstrated by Super-Kamiokande, the two events point to the cells of maximum overburden, where only a neutrino-induced signal survives.

Conversely, the neutrino hypothesis predicts identifiable events precisely in the observed charge window and in the highest-overburden cells, both observed, with an emerging lepton above about 20~TeV required by the candidate charges and no upper bound available from the published calibration. Under the conventional atmospheric flux, the most probable interpretation of each candidate is a $\nu_\mu$-induced muon: the atmospheric $\nu_\tau$ component is negligible, and $\nu_\mu \to \nu_\tau$ oscillations are inoperative over these baselines at TeV energies; the $\nu_\tau \to \tau$ channel becomes relevant only through the astrophysical multi-PeV flux suggested by the $\sim$2.1--2.5$\sigma$ rate excess.

Two previously published expectations are not identical. Ref.~\cite{Albert2022} anticipates that, with no improvement to the existing algorithms, at least one neutrino-induced muon should be detected every couple of years on average; that figure is for muons above 100~GeV over its full analysis region, with the effective area increased by at least two orders of magnitude at high energy, and it is combined with the neutrino-induced muon intensity of Ref.~\cite{Crouch1978}. It is not an expectation for the identifiable large-charge events discussed here, and Ref.~\cite{Albert2022} makes no statement about the two candidates: its analysis region, chosen to optimise reconstruction efficiency for the characterisation of the scattered muon background, does not include the azimuth of cell E. The expectation relevant to this work is instead the $\lambda_{\rm tot} \approx 0.11$--$0.2$ of Section~\ref{sec:rates}, which applies to the same selection that produced the candidates.

Measured against that expectation, two events are anomalous but of low statistical significance, and at least one of the two deposits is extraordinary: $1744.8~\pe$ per pixel is a factor $\approx 21$ above the observed scattered-muon population and a factor $\approx 4.7$ above the brightest muon in the published simulation, which no scattering scenario reproduces. The origin of the excess is therefore an open question, and with two events a statistical fluctuation remains the most probable explanation. HAWC has completed more than ten years of operation \cite{HAWC10years}, so the accumulated livetime exceeds the 181 published days by more than an order of magnitude. The 1.5-year sample of Ref.~\cite{Castellon2025} already suffices: it is three times the published livetime, so the conventional expectation is $0.3$--$0.6$ events whereas the observed rate would yield $\sim$6, a separation of $3.9$--$4.7\sigma$ with a $72$--$85\%$ probability of exceeding $3\sigma$ if the excess is real; and if it was a fluctuation, the most likely outcome by far is no additional event at all, which by itself would settle the question. If the present interpretation is correct, these constitute the first neutrino candidates detected in Mexico and the first obtained with the Earth-skimming technique using a volcano as the neutrino target.

\section*{CRediT authorship contribution statement}

\textbf{Hermes Le\'on Vargas:} Conceptualization, Methodology, Formal analysis, Investigation, Visualization, Funding acquisition, Writing ---
original draft, Writing --- review \& editing.
\textbf{Andr\'es Sandoval:} Conceptualization, Validation, Funding
acquisition, Writing --- review \& editing.

\section*{Declaration of competing interest}

The authors declare that they have no known competing financial interests or personal relationships that could have appeared to influence the work reported in this paper.

\section*{Acknowledgements}

Research carried out thanks to the Program for Support of Research and Technological Innovation Projects (PAPIIT) of UNAM, grant IN100227. We appreciate the support received from the UNAM Physics Institute Research Program 2026 Call (PIIF26).

\section*{Data availability}

No new experimental data were generated for this study. All numerical inputs are published values, compiled in Table~\ref{tab:inputs} together with their sources. The digital elevation model used for the independent overburden validation is publicly available from INEGI (Continuo de Elevaciones Mexicano, CEM 3.0).

\section*{Declaration of generative AI and AI-assisted technologies in the manuscript preparation process}

During the preparation of this work the authors used Claude (Anthropic) in order to verify the numerical and internal consistency of the manuscript and to assist with editing and formatting. After using this tool, the authors reviewed and edited the content as needed and take full responsibility for the content of the published article.

\bibliographystyle{elsarticle-num} \bibliography{hawc_neutrinos}

\begin{thebibliography}{10}
\expandafter\ifx\csname url\endcsname\relax
  \def\url#1{\texttt{#1}}\fi
\expandafter\ifx\csname urlprefix\endcsname\relax\def\urlprefix{URL }\fi
\expandafter\ifx\csname href\endcsname\relax
  \def\href#1#2{#2} \def\path#1{#1}\fi

\bibitem{KGF1971}
M.~R. Krishnaswamy, M.~G.~K. Menon, V.~S. Narasimham, K.~Hinotani, N.~Ito,
  S.~Miyake, J.~L. Osborne, A.~J. Parsons, A.~W. Wolfendale, The {Kolar Gold
  Fields} neutrino experiment {I}, Proc. R. Soc. Lond. A 323 (1971) 489.

\bibitem{KGF1975}
M.~R. Krishnaswamy, M.~G.~K. Menon, V.~S. Narasimham, N.~Ito, S.~Kawakami,
  S.~Miyake, Evidence for the production of new particles in cosmic ray
  experiments deep underground, Pramana 5 (1975) 59.
\newblock \href {https://doi.org/10.1007/BF02846033}
  {\path{doi:10.1007/BF02846033}}.

\bibitem{Achar1965}
C.~V. Achar, et~al., Detection of muons produced by cosmic ray neutrinos deep
  underground, Phys. Lett. 18 (1965) 196.

\bibitem{Reines1965}
F.~Reines, et~al., Evidence for high-energy cosmic-ray neutrino interactions,
  Phys. Rev. Lett. 15 (1965) 429.

\bibitem{Frejus1996}
W.~Rhode, et~al., Limits on the flux of very high energy neutrinos with the
  {Fr{\'e}jus} detector, Astropart. Phys. 4 (1996) 217, {Fr{\'e}jus
  Collaboration}.

\bibitem{MACRO2003}
M.~Ambrosio, et~al., Search for diffuse neutrino flux from astrophysical
  sources with {MACRO}, Astropart. Phys. 19 (2003) 1, {MACRO Collaboration}.

\bibitem{Soudan2_1999}
W.~W.~M. Allison, et~al., The atmospheric neutrino flavor ratio from a 3.9
  fiducial kiloton-year exposure of {Soudan 2}, Phys. Lett. B 449 (1999) 137,
  {Soudan 2 Collaboration}.

\bibitem{LVD1995}
M.~Aglietta, et~al., Neutrino-induced and atmospheric single-muon fluxes
  measured over five decades of intensity by {LVD} at {Gran Sasso Laboratory},
  Astropart. Phys. 3 (1995) 311, {LVD Collaboration}.

\bibitem{IC40}
R.~Abbasi, et~al., Measurement of the atmospheric neutrino energy spectrum from
  100 {GeV} to 400 {TeV} with {IceCube}, Phys. Rev. D 83 (2011) 012001,
  {IceCube Collaboration}.

\bibitem{AMANDA2010}
R.~Abbasi, et~al., The energy spectrum of atmospheric neutrinos between 2 and
  200 {TeV} with the {AMANDA-II} detector, Astropart. Phys. 34 (2010) 48,
  {IceCube Collaboration}.

\bibitem{KM3NeT2025}
S.~Aiello, et~al., Observation of an ultra-high-energy cosmic neutrino with
  {KM3NeT}, Nature 638 (2025) 376, {KM3NeT Collaboration}.

\bibitem{SnowmassNF10}
J.~R. Klein, et~al., {Snowmass} neutrino frontier {NF10} topical group report:
  Neutrino detectors (2022).
\newblock \href {http://arxiv.org/abs/2211.09669} {\path{arXiv:2211.09669}}.

\bibitem{SnowmassHE}
M.~Ackermann, et~al., High-energy and ultra-high-energy neutrinos: A {Snowmass}
  white paper (2022).
\newblock \href {http://arxiv.org/abs/2203.08096} {\path{arXiv:2203.08096}}.

\bibitem{TAMBO2026}
C.~A. Arg{\"u}elles, et~al., Measuring the high-energy neutrino sky using the
  deep-valley neutrino observatory {TAMBO}, Nat. Astron. 10 (2026) 947, {TAMBO
  Collaboration}.
\newblock \href {https://doi.org/10.1038/s41550-026-02916-4}
  {\path{doi:10.1038/s41550-026-02916-4}}.

\bibitem{Fargion1999}
D.~Fargion, A.~Aiello, R.~Conversano, Horizontal tau air showers from mountains
  in deep valley, in: Proc. 26th International Cosmic Ray Conference, Vol.~2,
  1999, p. 396.
\newblock \href {http://arxiv.org/abs/astro-ph/9906450}
  {\path{arXiv:astro-ph/9906450}}.

\bibitem{Feng2002}
J.~L. Feng, P.~Fisher, F.~Wilczek, T.~M. Yu, Observability of earth-skimming
  ultrahigh energy neutrinos, Phys. Rev. Lett. 88 (2002) 161102.

\bibitem{Auger2009}
J.~Abraham, et~al., Limit on the diffuse flux of ultrahigh energy tau neutrinos
  with the surface detector of the {Pierre Auger Observatory}, Phys. Rev. D 79
  (2009) 102001, {Pierre Auger Collaboration}.

\bibitem{MAGIC2018}
M.~L. Ahnen, et~al., Limits on the flux of tau neutrinos from 1 {PeV} to 3
  {EeV} with the {MAGIC} telescopes, Astropart. Phys. 102 (2018) 77, {MAGIC
  Collaboration}.

\bibitem{Ashra2013}
Y.~Asaoka, M.~Sasaki, Cherenkov $\tau$ shower earth-skimming method for
  {PeV}--{EeV} $\nu_\tau$ observation with {Ashra}, Astropart. Phys. 41 (2013)
  7.

\bibitem{LeonVargas2017}
H.~Le{\'o}n~Vargas, A.~Sandoval, E.~Belmont, R.~Alfaro, Capability of the
  {HAWC} gamma-ray observatory for the indirect detection of ultrahigh-energy
  neutrinos, Adv. Astron. 2017 (2017) 1932413.

\bibitem{Albert2022}
A.~Albert, et~al., Characterization of the background for a neutrino search
  with the {HAWC} observatory, Astropart. Phys. 137 (2022) 102670, {HAWC
  Collaboration}.

\bibitem{Smith2015}
A.~J. Smith, {HAWC}: design, operation, reconstruction and analysis, in: Proc.
  34th International Cosmic Ray Conference, Vol. ICRC2015 of PoS, 2015, p. 966,
  for the HAWC Collaboration.
\newblock \href {http://arxiv.org/abs/1508.05826} {\path{arXiv:1508.05826}}.

\bibitem{ICRC2019}
H.~Le{\'o}n~Vargas, Prospects of earth-skimming neutrino detection with {HAWC},
  in: Proc. 36th International Cosmic Ray Conference, Vol. ICRC2019 of PoS,
  2019, p. 940, for the HAWC Collaboration.

\bibitem{Volkova1980}
L.~V. Volkova, Energy spectra and angular distributions of atmospheric
  neutrinos, Sov. J. Nucl. Phys. 31 (1980) 784--790, [Yad. Fiz. \textbf{31}
  (1980) 1510].

\bibitem{GaisserHonda2002}
T.~K. Gaisser, M.~Honda, Flux of atmospheric neutrinos, Ann. Rev. Nucl. Part.
  Sci. 52 (2002) 153--199.
\newblock \href {https://doi.org/10.1146/annurev.nucl.52.050102.090645}
  {\path{doi:10.1146/annurev.nucl.52.050102.090645}}.

\bibitem{IceCubeTG2022}
R.~Abbasi, et~al., Improved characterization of the astrophysical muon-neutrino
  flux with 9.5 years of icecube data, Astrophys. J. 928~(1) (2022) 50.
\newblock \href {http://arxiv.org/abs/2111.10299} {\path{arXiv:2111.10299}},
  \href {https://doi.org/10.3847/1538-4357/ac4d29}
  {\path{doi:10.3847/1538-4357/ac4d29}}.

\bibitem{IceCubeST2024}
R.~Abbasi, et~al., Characterization of the astrophysical diffuse neutrino flux
  using starting track events in icecube, Phys. Rev. D 110~(2) (2024) 022001.
\newblock \href {http://arxiv.org/abs/2402.18026} {\path{arXiv:2402.18026}},
  \href {https://doi.org/10.1103/PhysRevD.110.022001}
  {\path{doi:10.1103/PhysRevD.110.022001}}.

\bibitem{HAWCDetector2023}
A.~U. Abeysekara, et~al., The {High-Altitude Water Cherenkov} ({HAWC})
  observatory in {M{\'e}xico}: the primary detector, Nucl. Instrum. Meth. A
  1052 (2023) 168253, {HAWC Collaboration}.
\newblock \href {http://arxiv.org/abs/2304.00730} {\path{arXiv:2304.00730}}.

\bibitem{Lohmann1985}
W.~Lohmann, R.~Kopp, R.~Voss, Energy loss of muons in the energy range 1--10000
  {GeV}, CERN Yellow Report CERN-85-03, CERN (1985).

\bibitem{Groom2001}
D.~E. Groom, N.~V. Mokhov, S.~I. Striganov, Muon stopping power and range
  tables 10 {MeV}--100 {TeV}, At. Data Nucl. Data Tables 78 (2001) 183.

\bibitem{INEGI}
{INEGI}, Continuo de elevaciones mexicano 3.0 ({CEM} 3.0),
  \url{https://www.inegi.org.mx/app/geo2/elevacionesmex/}, accessed August 2026
  (2013).

\bibitem{Barrett1952}
P.~H. Barrett, L.~M. Bollinger, G.~Cocconi, Y.~Eisenberg, K.~Greisen,
  Interpretation of cosmic-ray measurements far underground, Rev. Mod. Phys. 24
  (1952) 133.

\bibitem{MenonMurthy1967}
M.~G.~K. Menon, P.~V. Ramana~Murthy, Cosmic ray intensities deep underground,
  Prog. Elem. Part. Cosmic Ray Phys. 9 (1967) 161.

\bibitem{PDG2020}
P.~A. Zyla, et~al., Review of particle physics, Prog. Theor. Exp. Phys. 2020
  (2020) 083C01, ch. 34: Passage of Particles Through Matter.

\bibitem{Sternheimer1984}
R.~M. Sternheimer, M.~J. Berger, S.~M. Seltzer, Density effect for the
  ionization loss of charged particles in various substances, At. Data Nucl.
  Data Tables 30 (1984) 261.

\bibitem{NuLeptonSim2024}
A.~Cummings, R.~Krebs, S.~Wissel, J.~Alvarez-Mu{\~n}iz, W.~R. Carvalho~Jr.,
  A.~Romero-Wolf, H.~Schoorlemmer, E.~Zas, Secondary lepton production,
  propagation, and interactions with {NuLeptonSim} (2023).
\newblock \href {http://arxiv.org/abs/2311.03646} {\path{arXiv:2311.03646}}.

\bibitem{IceCubeMuons2016}
M.~G. Aartsen, et~al., Characterization of the atmospheric muon flux in
  {IceCube}, Astropart. Phys. 78 (2016) 1, {IceCube Collaboration}.

\bibitem{Matsuno1984}
S.~Matsuno, et~al., Cosmic-ray muon spectrum up to 20 {TeV} at 89$^\circ$
  zenith angle, Phys. Rev. D 29 (1984) 1.

\bibitem{ChirkinMMC}
D.~Chirkin, W.~Rhode, Propagating leptons through matter with {Muon Monte
  Carlo} ({MMC}) (2004).
\newblock \href {http://arxiv.org/abs/hep-ph/0407075}
  {\path{arXiv:hep-ph/0407075}}.

\bibitem{Elbert1983}
J.~W. Elbert, T.~K. Gaisser, T.~Stanev, Analysis of deep-underground muons,
  Phys. Rev. D 27 (1983) 1448.

\bibitem{Inazawa1983}
H.~Inazawa, K.~Kobayakawa, Production of prompt cosmic ray muons and neutrinos,
  Prog. Theor. Phys. 69 (1983) 1195.

\bibitem{SuperK2005}
Y.~Ashie, et~al., Measurement of atmospheric neutrino oscillation parameters by
  {Super-Kamiokande I}, Phys. Rev. D 71 (2005) 112005, {Super-Kamiokande
  Collaboration}.

\bibitem{Ave2000}
M.~Ave, R.~A. V\'azquez, E.~Zas, Modelling horizontal air showers induced by
  cosmic rays, Astropart. Phys. 14 (2000) 91.
\newblock \href {http://arxiv.org/abs/astro-ph/0011490}
  {\path{arXiv:astro-ph/0011490}}.

\bibitem{Sako2009}
T.~K. Sako, et~al., Exploration of a 100 {TeV} gamma-ray northern sky using the
  {Tibet} air-shower array combined with an underground water-{Cherenkov}
  muon-detector array, Astropart. Phys. 32 (2009) 177.

\bibitem{TALE2018}
R.~U. Abbasi, et~al., The cosmic-ray energy spectrum between 2 {PeV} and 2
  {EeV} observed with the {TALE} detector in monocular mode, Astrophys. J. 865
  (2018) 74, {Telescope Array Collaboration}.

\bibitem{Gandhi1998}
R.~Gandhi, C.~Quigg, M.~H. Reno, I.~Sarcevic, Neutrino interactions at
  ultrahigh energies, Phys. Rev. D 58 (1998) 093009.

\bibitem{LipariStanev1991}
P.~Lipari, T.~Stanev, Propagation of multi-{TeV} muons, Phys. Rev. D 44 (1991)
  3543.

\bibitem{BGR18}
V.~Bertone, R.~Gauld, J.~Rojo, Neutrino telescopes as {QCD} microscopes, JHEP
  01 (2019) 217.
\newblock \href {http://arxiv.org/abs/1808.02034} {\path{arXiv:1808.02034}}.

\bibitem{Kistler2016}
M.~D. Kistler, R.~Laha, Multi-{PeV} signals from a new astrophysical neutrino
  flux beyond the {Glashow} resonance, Phys. Rev. Lett. 120 (2018) 241105.
\newblock \href {http://arxiv.org/abs/1605.08781} {\path{arXiv:1605.08781}},
  \href {https://doi.org/10.1103/PhysRevLett.120.241105}
  {\path{doi:10.1103/PhysRevLett.120.241105}}.

\bibitem{PUEO2021}
Q.~Abarr, et~al., The {Payload for Ultrahigh Energy Observations} ({PUEO}): a
  white paper, JINST 16 (2021) P08035, {PUEO Collaboration}.
\newblock \href {http://arxiv.org/abs/2010.02892} {\path{arXiv:2010.02892}}.

\bibitem{Palmisano2026}
S.~Palmisano, D.~Redigolo, M.~Tammaro, A.~Tesi, The soft volume of ultra-high
  energy neutrinos experiments (2026).
\newblock \href {http://arxiv.org/abs/2607.13143} {\path{arXiv:2607.13143}}.

\bibitem{Chattopadhyay2026}
D.~S. Chattopadhyay, C.~A. Arg{\"u}elles, V.~Brdar, Four, one, and none:
  Quantifying the ultra-high-energy neutrino anomaly across {ANITA-IV},
  {KM3NeT}, and {IceCube} (2026).
\newblock \href {http://arxiv.org/abs/2607.19487} {\path{arXiv:2607.19487}}.

\bibitem{HAWCCalib2015}
H.~A. Ayala~Solares, M.~Gerhardt, C.~M. Hui, R.~J. Lauer, Z.~Ren,
  F.~Salesa~Greus, H.~Zhou, The calibration system of the {HAWC} gamma-ray
  observatory, in: Proc. 34th International Cosmic Ray Conference, Vol.
  ICRC2015 of PoS, 2015, p. 997, for the HAWC Collaboration.
\newblock \href {http://arxiv.org/abs/1508.04312} {\path{arXiv:1508.04312}}.

\bibitem{Castellon2025}
C.~Castell{\'o}n, H.~Le{\'o}n~Vargas, Search for seasonal variations of the
  horizontal muon rate with the {HAWC} observatory, in: Proc. 39th
  International Cosmic Ray Conference, Vol. ICRC2025 of PoS, 2025, p. 316, for
  the HAWC Collaboration.

\bibitem{AngelesCamacho2021}
J.~R. Angeles~Camacho, H.~Le{\'o}n~Vargas, Horizontal muon track identification
  with neural networks in {HAWC}, in: Proc. 37th International Cosmic Ray
  Conference, Vol. ICRC2021 of PoS, 2021, p. 1036, for the HAWC Collaboration.

\bibitem{Crouch1978}
M.~F. Crouch, P.~B. Landecker, J.~F. Lathrop, F.~Reines, W.~G. Sandie, H.~W.
  Sobel, H.~Coxell, J.~P.~F. Sellschop, Cosmic ray muon fluxes deep
  underground: Intensity versus depth, and the neutrino induced component,
  Phys. Rev. D 18 (1978) 2239.
\newblock \href {https://doi.org/10.1103/PhysRevD.18.2239}
  {\path{doi:10.1103/PhysRevD.18.2239}}.

\bibitem{HAWC10years}
{HAWC Collaboration}, The first 10 years of the {HAWC} gamma-ray observatory:
  science results, Rev. Mex. Astron. Astrofis. 61 (2025) 261.
\newblock \href {https://doi.org/10.22201/ia.01851101p.2025.61.03.13}
  {\path{doi:10.22201/ia.01851101p.2025.61.03.13}}.

\end{thebibliography}

\end{document}